\documentclass[10pt]{article}
\usepackage{amssymb,amsmath}
\usepackage{amsthm}
\usepackage{mathrsfs}  
\usepackage{graphicx}
\usepackage{dcolumn}
\usepackage{bm}
\usepackage{fullpage}
\usepackage{color}
\usepackage[all]{xy}
\begin{document}
\newtheorem{Def}{Definition}[section]
\newtheorem{Thm}{Theorem}[section]
\newtheorem{Proposition}{Proposition}[section] 
\newtheorem{Lemma}{Lemma}[section]
\theoremstyle{definition}
\newtheorem*{Proof}{Proof}
\newtheorem{Example}{Example}[section] 
\newtheorem{Postulate}{Postulate}[section]
\newtheorem{Corollary}{Corollary}[section]
\newtheorem{Remark}{Remark}[section]
\theoremstyle{remark}
\newcommand{\beq}{\begin{equation}}
\newcommand{\beqa}{\begin{eqnarray}}
\newcommand{\eeq}{\end{equation}}
\newcommand{\eeqa}{\end{eqnarray}}
\newcommand{\non}{\nonumber}
\newcommand{\fr}[1]{(\ref{#1})}
\newcommand{\const}{\mbox{Const.}}
\newcommand{\bb}{\mbox{\boldmath {$b$}}}
\newcommand{\bbe}{\mbox{\boldmath {$e$}}}
\newcommand{\bt}{\mbox{\boldmath {$t$}}}
\newcommand{\bn}{\mbox{\boldmath {$n$}}}
\newcommand{\br}{\mbox{\boldmath {$r$}}}
\newcommand{\bC}{\mbox{\boldmath {$C$}}}
\newcommand{\bH}{\mbox{\boldmath {$H$}}}
\newcommand{\bp}{\mbox{\boldmath {$p$}}}
\newcommand{\bx}{\mbox{\boldmath {$x$}}}
\newcommand{\bF}{\mbox{\boldmath {$F$}}}
\newcommand{\bT}{\mbox{\boldmath {$T$}}}
\newcommand{\bomega}{\mbox{\boldmath {$\omega$}}}
\newcommand{\ve}{{\varepsilon}}
\newcommand{\e}{\mathrm{e}}
\newcommand{\F}{\mathrm{F}}
\newcommand{\Loc}{\mathrm{Loc}}
\newcommand{\Ree}{\mathrm{Re}}
\newcommand{\Imm}{\mathrm{Im}}
\newcommand{\Hess}{\mathrm{Hess}}
\newcommand{\hF}{\widehat F}
\newcommand{\hL}{\widehat L}
\newcommand{\tA}{\widetilde A}
\newcommand{\tB}{\widetilde B}
\newcommand{\tC}{\widetilde C}
\newcommand{\tL}{\widetilde L}
\newcommand{\tK}{\widetilde K}
\newcommand{\tX}{\widetilde X}
\newcommand{\tY}{\widetilde Y}
\newcommand{\tU}{\widetilde U}
\newcommand{\tZ}{\widetilde Z}
\newcommand{\talpha}{\widetilde \alpha}
\newcommand{\te}{\widetilde e}
\newcommand{\tv}{\widetilde v}
\newcommand{\ts}{\widetilde s}
\newcommand{\tx}{\widetilde x}
\newcommand{\ty}{\widetilde y}
\newcommand{\ud}{\underline{\delta}}
\newcommand{\uD}{\underline{\Delta}}
\newcommand{\chN}{\check{N}}
\newcommand{\cA}{{\cal A}}
\newcommand{\cB}{{\cal B}}
\newcommand{\cC}{{\cal C}}
\newcommand{\cD}{{\cal D}}
\newcommand{\cE}{{\cal E}}
\newcommand{\cF}{{\cal F}}
\newcommand{\cH}{{\cal H}}
\newcommand{\cI}{{\cal I}}
\newcommand{\cL}{{\cal L}}
\newcommand{\cM}{{\cal M}}
\newcommand{\cN}{{\cal N}}
\newcommand{\cR}{{\cal R}}
\newcommand{\cS}{{\cal S}}
\newcommand{\cY}{{\cal Y}}
\newcommand{\cZ}{{\cal Z}}
\newcommand{\cU}{{\cal U}}
\newcommand{\cV}{{\cal V}}
\newcommand{\tcA}{\widetilde{\cal A}}
\newcommand{\DD}{{\cal D}}
\newcommand\TYPE[3]{ \underset {(#1)}{\overset{{#3}}{#2}}  }
\newcommand{\bfe}{\boldsymbol e} 
\newcommand{\bfb}{{\boldsymbol b}}
\newcommand{\bfd}{{\boldsymbol d}}
\newcommand{\bfh}{{\boldsymbol h}}
\newcommand{\bfj}{{\boldsymbol j}}
\newcommand{\bfn}{{\boldsymbol n}}
\newcommand{\bfA}{{\boldsymbol A}}
\newcommand{\bfB}{{\boldsymbol B}}
\newcommand{\bfJ}{{\boldsymbol J}}
\newcommand{\dr}{\mathrm{d}}
\newcommand{\TE}{\mathrm{TE}}
\newcommand{\TM}{\mathrm{TM}}
\newcommand{\Ai}{\mathrm{Ai}}
\newcommand{\Bi}{\mathrm{Bi}}
\newcommand{\sech}{\mathrm{sech}}
\newcommand{\jthree}{  \TYPE 3  {j}  {}   }
\newcommand{\Lam}{ \TYPE q  {\Lambda}   {}   }
\newcommand{\alp}{ \TYPE p  {\alpha}   {}   }
\newcommand{\al}[1]{ \TYPE {#1}  {\alpha}   {}   }
\newcommand{\bep}{ \TYPE p  {\beta}   {}   }
\newcommand{\be}[1]{ \TYPE {#1}  {\beta}   {}   }
\newcommand{\gamq}{ \TYPE q  {\gamma}   {}   }
\newcommand{\hash}{\#}
\newcommand{\hashat}{\widehat{\#}}
\newcommand{\hashch}{\stackrel{\vee}{\#}}
\newcommand{\chd}{\stackrel{\vee}{\D}}
\newcommand\NN[1]{{\cal N}_{#1}}
\newcommand\MM[1]{{\cal M}_{#1}}
\newcommand\BAE[1]{{\begin{equation}{\begin{aligned}#1\end{aligned}}\end{equation}}}
\newcommand{\GamCLamM}[1]{{\Gamma\mathbb{C}\Lambda^{{#1}}\cal{M}}}
\newcommand{\GamLamM}[1]{{\Gamma\Lambda^{{#1}}\,\cal{M}}}
\newcommand{\GamLamU}[1]{{\Gamma\Lambda^{{#1}}\,\cal{U}}}
\newcommand{\GamLamHU}[1]{{\Gamma\Lambda^{{#1}}\,\widehat{\cal{U}}}}
\newcommand{\GamLam}[2]{{\Gamma\Lambda^{{#1}}\,{#2}}}
\newcommand{\GTM}{{\Gamma T\cal{M}}}
\newcommand{\GTU}{{\Gamma T\cal{U}}}
\newcommand{\GT}[1]{{\Gamma T {#1}}}
\newcommand{\normM}[2]{\left(  #1\, , \, #2 \right)}
\newcommand{\normU}[2]{\left\{ #1\, , \, #2 \right\}}
\newcommand{\diag}[1]{\mbox{diag}\{\, #1\,\}}
\newcommand{\GtM}[2]{\Gamma T^{#1}_{#2}{\cal M}}
\newcommand{\inp}[2]{\left\langle\,  #1\, , \, #2\, \right\rangle}
\newcommand{\defi}{\noindent {\bf Definition : } }
\newcommand{\prop}{\noindent {\bf Proposition : } }
\newcommand{\theo}{\noindent {\bf Theorem : } }
\newcommand{\exam}{\noindent {\bf Example : } }
\newcommand{\equp}[1]{\overset{\mathrm{#1}}{=}}
\newcommand{\wt}[1]{\widetilde{#1}}
\newcommand{\wh}[1]{\widehat{#1}}
\newcommand{\ch}[1]{\check{#1}}
\newcommand{\ii}{\imath}
\newcommand{\ic}{\iota}
\newcommand{\mi}{\,\mathrm{i}\,}
\newcommand{\mr}{\,\mathrm{r}\,}
\newcommand{\mbbC}{\mathbb{C}}
\newcommand{\mbbD}{\mathbb{D}}
\newcommand{\mbbR}{\mathbb{R}}
\newcommand{\mbbZ}{\mathbb{Z}}
\newcommand{\Leftrightup}[1]{\overset{\mathrm{#1}}{\Longleftrightarrow}}
\newcommand{\ol}[1]{\overline{#1}}
\newcommand{\rmC}{\mathrm{C}}
\newcommand{\rmH}{\mathrm{H}}
\newcommand{\Id}{\mathrm{Id}} 
\title{\bf 
Hessian-information geometric formulation\\ 
of Hamiltonian systems and 
generalized\\ Toda's dual transform
}
\author{  {\bf Shin-itiro Goto}$^\dag$ {\bf and  Tatsuaki Wada}$^\ddag$,\\ 
$\dag$Department of Applied Mathematics and Physics, 
Graduate School of Informatics, \\
Kyoto University, Yoshida Honmachi, Sakyo-ku, Kyoto, 606-8501, Japan\\
$\ddag$
Region of Electrical and Electronic Systems Engineering, \\ 
Ibaraki University, Nakanarusawacho, Hitachi 316-8511, Japan
} 
\date{\hfill}
\maketitle
\begin{abstract}%
In this paper a class of classical Hamiltonian systems is geometrically 
formulated. 
This class is such that a Hamiltonian 
can be written as the sum of a kinetic energy function and a 
potential energy function. 
In addition, these energy functions are assumed strictly convex.
For this class of Hamiltonian systems Hessian and information 
geometric formulation is given.
With this formulation, a generalized Toda's dual transform is proposed, 
where his original 
transform was used in deriving his integrable lattice system. 
Then a relation between the generalized Toda's dual transform and 
the Legendre transform of a class of potential energy functions is shown. 
As an extension of this formulation, 
dissipation-less electric circuit models are also discussed in the 
geometric viewpoint above.  
\end{abstract}%
\section{Introduction}
Ideas of duality shed light on various aspects in mathematics and physics. 
One of such is the Legendre duality 
by the Legendre transform
associated with a convex function, and  
this transform connects different viewpoints.
 
In mathematics, the Legendre transform plays a  role in 
Hessian geometry\,\cite{Shima}.  
This geometry is deeply connected to information geometry, where 
information geometry is a geometrization of mathematical 
statistics\,\cite{AN,Ay2017}.  
Applications of information geometry include 
statistical interference, thermodynamics, and so on.  
In these days 
some of Hessian geometry and that of 
information geometry are  
amalgamated\,\cite{Matsuzoe2013,Furuhata2009}. 
Then, it is expected that 
the development of Hessian and information 
geometries influences various mathematical sciences. 
The Legendre transform also appears in contact geometry, where
contact geometry is known to be an odd-dimensional cousin of 
symplectic geometry\,\cite{Arnold,Silva2008}.  
As is well-known, symplectic geometry is  
a geometrization of analytical mechanics, and 
contact and symplectic geometries have influenced various branches 
of pure and applied mathematics. Such 
examples include geometric formulations of electric circuit 
models\,\cite{Goto2016,Eberard2006} and a geometric characterization for 
dynamical systems discussed in information geometry\,\cite{BN2016}.  

In physics, the subjects where the use of the Legendre transform 
is emphasized are  
analytical mechanics and thermodynamics\,\cite{Goldstein,Callen}.   
Since analytical mechanics is 
placed at the center of theoretical physics,  
its development influences various physical sciences,
such as condensed matter physics, high energy physics, and so on. 
Thermodynamics is related to black hole physics,  
information geometry  and so on\,\cite{Quevedo2008, Goto2015,Wada2015}. 
Although the use of the Legendre transform has been stressed in 
analytical mechanics\,\cite{Teruel2013}, the emphasis is not placed on 
recent development of Hessian geometry and information one.
Also, the application of the Legendre transform has mainly been 
to kinetic energy functions.
We then feel that the link between geometry consisting of 
Hessian and information geometries and 
analytical mechanics or symplectic geometry should be explored 
more\,\cite{LZ2017}.
In addition, in Toda's paper of 1967 \cite{Toda1967}, the way to
find his integrable lattice model is to apply a transform, called the 
dual transform.
His dual transform makes nonlinear force terms linear ones for 
Hamilton's equations of motion. On the other hand the applications of 
the Legendre transform to a class of potential functions yield the   
linearized forces as well. Thus we should explore a relation between 
these transforms. If such a relation exists, it is then 
expected that this relation can be 
introduced in the theory of dissipation-less electric circuit models 
since model equations are similar to Hamilton's equations.

In this paper 
the term Hessian-information geometry stands for 
the geometry consisting of Hessian geometry and information one, and 
it is shown how the Legendre transform is 
applied to a class of classical Hamiltonian systems with emphasis on 
Hessian-information geometry. 
Here such a Hamiltonian is  
the sum of a strictly convex kinetic energy functions and strictly convex 
potential energy function, so that the Legendre transform is invertible.
In this Hessian-information geometric formulation, it is shown that 
the $\alpha$-connection invented in information geometry also appears 
in canonical equations of motion. 
Then Toda's dual transform and a class of lattice system are 
interpreted from the viewpoint of the 
Legendre transform and that of Hessian-information geometry.
Also, some dual lattice systems are constructed explicitly. 
Finally the Hessian-information geometric formulation for  
dissipation-less electric circuit theory is discussed. 
In appendix, a brief explanation of information geometry and 
Hessian geometry is given. 

\section{Natural Hamiltonian systems with strictly convex energies}
In this paper a class of Hamiltonian systems is considered. In this class 
a Hamiltonian is  
the sum of a kinetic energy function and a potential energy function. 

Let $\cM$ be a $2n$-dimensional manifold, $H:\cM\to\mbbR$ 
a Hamiltonian, $(p,q)$ a set of canonical coordinates such that 
a symplectic $2$-form is 
$\omega=\sum_{a=1}^{\,n}\dr p_{\,a}\wedge \dr q^{\,a}=\dr p_{\,a}\wedge \dr q^{\,a}$, 
$K:\cM\to\mbbR$ a kinetic energy 
function depending on  $p$ only, 
and $U:\cM\to\mbbR$ a potential energy function 
depending on  $q$ only. 
Here and in what follows the Einstein convention is used and 
every object is differentiable.  
Thus, $H$ can be written as    
\beq
H(q,p)
=K(p)+U(q).
\label{natural-Hamiltonian-general}
\eeq
The canonical equations of motion are then 
\beq
\frac{\dr}{\dr t}q^{\,a}
=\frac{\partial K}{\partial p_{\,a}},\qquad\mbox{and}\qquad 
\frac{\dr}{\dr t}p_{\,a}
=-\,\frac{\partial U}{\partial q_{\,a}},\qquad a\in\{1,\ldots,n\},
\label{canonical-equations-general}
\eeq 
where $t\in\mbbR$ denotes time. 
The equation \fr{canonical-equations-general}  
also referred to as Hamilton's equations of motion. These 
can be derived from 
$$
\ii_{X_{H}}\omega 
=-\,\dr H,
$$
where $X_{H}\in T\cM$ is a Hamiltonian vector field, and 
$\ii_{Y}$ the interior product operator with $Y\in T\cM$, with $T\cM$ being 
the tangent bundle. 
In this geometric context the triplet $(\cM,\omega,H)$ is referred  to as 
a (classical) Hamiltonian system.

If a Hamiltonian can be written as 
\fr{natural-Hamiltonian-general}, then $H$ is 
referred to as a natural Hamiltonian and its system is referred to 
as a natural Hamiltonian system.
Throughout this section it is assumed that  
\begin{itemize}
\item
the manifold $\cM$ can be written as $\cM=\cM_{\,K}\times\cM_{\,U}$ with 
some $n$-dimensional manifolds $\cM_{\,K}$ and $\cM_{\,U}$.  
Local coordinates of $\cM_{\,K}$ and $\cM_{\,U}$ 
are denoted as $p$ and $q$, respectively.
\end{itemize}

To specify $H$ given in \fr{natural-Hamiltonian-general}  further, 
strictly convex function is introduced. 
Let $\cN$ be an $n$-dimensional manifold, $\{x^{\,a}\}$ 
a set of coordinates, and $f:\cN\to\mbbR$ a function.
If $f$ satisfies  
$$
\left(\,\frac{\partial^2\, f}{\partial x^{\,a}\partial x^{\,b}}\right)
\succ 0,
$$
in some convex domain $\cD\subset\cN$,  
then $f$ is referred to as a strictly convex function in $\cD$.  
Here $A \succ 0$ denotes that a matrix $A$ is positive definite.

Then, strictly convex energy functions are introduced.
\begin{Def}
{\bf (Strictly convex energy functions).}\ 
If $K$ and $U$ satisfy
$$
\left(
\frac{\partial^2K}{\partial p_{\,a}\partial p_{\,b}}\right)\succ 0,\qquad 
\mbox{and}\qquad
\left(
\frac{\partial^2\,U}{\partial q^{\,a}\partial q^{\,b}}\right)\succ 0,\qquad 
a,b\in \{1,\ldots,n\}
$$ 
in some convex domains, then $K$ is referred to as 
a strictly convex kinetic energy function 
and $U$ a strictly convex potential energy function.
\end{Def}

If one considers a natural Hamiltonian system whose Hamiltonian 
is the sum of 
strictly convex energy functions, 
then one can apply Hessian geometry to the system.  Since a part of 
Hessian geometry has been applied to information geometry, one can also 
apply known facts found in information geometry to Hamiltonian systems with 
strictly convex energy functions.
\subsection{Non-vanishing potential systems}
\label{section-non-vanishing-potential-systems}
In this subsection it is assumed that 
\begin{itemize}
\item
a system is a natural Hamiltonian system whose Hamiltonian is the sum of 
strictly convex energy functions, $H=K+U$ with $K$ being a function of 
$p=\{\,p_{\,a}\,\}$,  
and $U$ being a function of $q=\{\,q^{\,a}\,\}$. 
\end{itemize}
From this assumption, 
the conditions 
$(\partial^{\,2}K/\partial p_{\,a}\partial p_{\,b})\succ 0$ and 
$(\partial^{\,2}\,U/\partial q^{\,a}\partial q^{\,b})\succ 0$,  
are satisfied. 

From convex analysis 
the following coordinates play various roles 
\begin{Def}
\label{definition-dual-coordinates-natural-Hamiltonian}
{\bf (Dual coordinates).}\, The coordinates  defined by 
$$
p_{\,*}^{\,a}
=\frac{\partial K}{\partial p_{\,a}},\qquad
 \mbox{and}\qquad
q_{\,a}^{\,*}
=\frac{\partial\, U}{\partial q^{\,a}},\qquad
a\in\{1,\ldots,n\}
$$
are referred to as dual coordinates. In particular, $p_{\,*}^{\,a}$ is 
referred to as being dual to $p_{\,a}$, and $q_{\,a}^{\,*}$ is referred to 
as being dual to $q^{\,a}$.
\end{Def}
\begin{Remark}
Since $K$ is strictly convex, one has that 
the correspondence between $p_{\,a}$ and $p_{\,*}^{\,a}$ is one-to-one. 
Similarly, the correspondence between 
$q^{\,a}$ and $q_{\,a}^{\,*}$ is also one-to-one.
\end{Remark}

Due to strict convexity of $K$ and $U$, one has the Riemannian 
metric tensor fields 
\beq
h^{\,K}=h_{\,K}^{\,ab}\,\dr p_{\,a}\otimes \dr p_{\,b},
\qquad\mbox{and}\qquad
h^{\,U}=h_{\,ab}^{\,U}\,\dr q^{\,a}\otimes \dr q^{\,b},
\label{metric-h}
\eeq
where 
\beq
h_{\,K}^{\,ab}
=\frac{\partial^2 K }{\partial p_{\,a}\partial p_{\,b}},\qquad\mbox{and}\qquad 
h_{\,ab}^{\,U}
=\frac{\partial^2\,U }{\partial q^{\,a}\partial q^{\,b}},
\qquad a,b\in\{1,\ldots,n\}.
\label{metric-h-components}
\eeq
\begin{Def}
{\bf (Riemannian metric tensor fields associated with convex energy functions).}\  
The $h^{\,K}$ in \fr{metric-h} with \fr{metric-h-components} 
is referred to as the Riemann metric tensor field associated with 
$K$, and $h^{\,U}$ in \fr{metric-h} with \fr{metric-h-components} is referred to 
as that associated with $U$, respectively. 
\end{Def}

There exist the inverse matrices of $(h_{\,K}^{\,ab})$ and $(h_{\,ab}^{\,U})$. 
Such inverse matrices $(h_{\,ab}^{\,K})$ and $(h_{\,U}^{\,ab})$  
can be written as   
$$
h_{\,ab}^{\,K}
=\frac{\partial^2 K^{\,*} }{\partial p_{\,*}^{\,a}\,\partial p_{\,*}^{\,b}},
\qquad\mbox{and}\qquad 
h_{\,U}^{\,ab}
=\frac{\partial^2\,U^{\,*} }{\partial q_{\,a}^{\,*}\,\partial q_{\,b}^{\,*}},
\qquad a,b\in\{1,\ldots,n\},
$$
where $K^{\,*}$ and  $U^{\,*}$ are the Legendre transforms of $K$ and $U$ : 
\beq
K^{\,*}(p_{\,*})
=\sup_{p}\left[\,p_{\,a}\,p_{\,*}^{\,a}-K(p)
\,\right],\qquad \mbox{and}\qquad
U^{\,*}(q^{\,*})
=\sup_{q}\left[\,q^{\,a}\,q_{\,a}^{\,*}-U(q)
\,\right].
\label{Legendre-transform-K-U}
\eeq
It can be shown that\,\cite{AN} 
\beq
p_{\,a}
=\frac{\partial K^{\,*}}{\partial p_{\,*}^{\,a} },\qquad\mbox{and}\qquad
q^{\,a}
=\frac{\partial\,U^{\,*}}{\partial q_{\,a}^{\,*} }.
\label{coordinates-by-dual-coordinates}
\eeq
The following inequalities are consequences of the 
strict convexity of $K$
and $U$.
\begin{Proposition}
\label{fact-divergence-K}
Let $z_{\,K}$ and $z_{\,K}^{\,\prime}$ be two points of $\cM_{\,K}$, 
$p=\{\,p_{\,a}\,\}$ and $p^{\,\prime}=\{\,p_{\,a}^{\,\prime}\,\}$ coordinates 
of $z_{\,K}$ and $z_{\,K}^{\,\prime}$, $p_{\,*}=\{\,p_{\,*}^{\,a}\,\}$ and 
$p_{\,*}^{\,\prime}=\{\,p_{\,*}^{\,\prime\, a}\,\}$ 
dual coordinates of $z_{\,K}$ and $z_{\,K}^{\,\prime}$, 
and $\mbbD_{\,K}:\cM_{\,K}\times\cM_{\,K}\to\mbbR$ 
a function such that
$$
\mbbD_{\,K}\, (\,z_{\,K}\,\|\,z_{\,K}^{\,\prime}\,)
=K(p)+K^{\,*}(p_{\,*}^{\,\prime})
-p_{\,a}\,p_{\,*}^{\,\prime\,a}.
$$
Then, it follows that 
$$
\mbbD_{\,K}\, (\,z_{\,K}\,\|\,z_{\,K}^{\,\prime}\,)
\geq 0.
$$
In addition, the equality holds when $z_{\,K}=z_{\,K}^{\,\prime}$.
\end{Proposition}
\begin{Proof}
See \cite{Shima} for example.
\qed
\end{Proof}
Similar to this, one has the following. 
\begin{Proposition}
\label{fact-divergence-U}
Let $z_{\,U}$ and $z_{\,U}^{\,\prime}$ be two points of $\cM_{\,U}$, 
$q=\{\,q^{\,a}\,\}$ and $q^{\,\prime}=\{\,q^{\,\prime\,a}\,\}$ coordinates of 
$z_{\,U}$ and $z_{\,U}^{\,\prime}$, $q^{\,*}=\{\,q_{\,a}^{\,*}\,\}$ and 
$q^{\,\prime\,*}=\{\,q_{\,a}^{\,\prime\, *}\,\}$ 
dual coordinates of $z_{\,U}$ and $z_{\,U}^{\,\prime}$, 
and $\mbbD_{\,U}:\cM_{\,U}\times\cM_{\,U}\to\mbbR$ 
a function such that
$$
\mbbD_{\,U}\, (\,z_{\,U}\,\|\,z_{\,U}^{\,\prime}\,)
=U(q)+U^{\,*}(\,q_{\,*}^{\,\prime}\,)
-q_{\,a}\,q_{\,*}^{\,\prime\,a}.
$$
Then, it follows that 
$$
\mbbD_{\,U}\, (\,z_{\,U}\,\|\,z_{\,U}^{\,\prime}\,)
\geq 0.
$$
In addition, the equality holds when $z_{\,U}=z_{\,U}^{\,\prime}$.
\end{Proposition}
\begin{Proof}
See \cite{Shima} for example.
\qed
\end{Proof}
In information geometry the functions similar to 
$\mbbD_{\,K}$ and $\mbbD_{\,U}$ in propositions\,\ref{fact-divergence-K} and 
\ref{fact-divergence-U} are often used. Such functions are known 
as the canonical divergences,  
and they are used in various applications\,
\cite{AN,Ay2017,Fujiwara1995}. 
As shown in these propositions, it should be 
emphasized that the canonical divergences can be introduced in 
the present class of Hamilton's equations, and that 
the existence of these functions enables one to discuss 
information geometric aspects of Hamiltonian systems.



Applying Hessian-information geometry to  
natural Hamiltonian systems, one can write 
canonical equations of motion in terms of geometric objects 
developed in such geometry. 
To this end, introducing some connections on Riemannian manifolds, 
one has Hessian manifolds. To discuss Hessian geometry of 
canonical equations of motion one defines 
the following connections. 
\begin{Def}
{\bf (Flat connections associated with energy functions).}\,
The connections $\nabla^{\,K}$ and $\nabla^{\,U}$ such that 
\beq
h^{\,K}=\nabla^{\,K}\dr K,\qquad\mbox{and}\qquad 
h^{\,U}=\nabla^{\,U}\dr U,
\label{definition-connections-K-U}
\eeq
are referred to as the connection associated with $K$, and referred to 
as that associated  with $U$, respectively. Also, 
the connections $\nabla^{\,K^{*}}$ and $\nabla^{\,U^{*}}$ such that 
\beq
h^{\,K}=\nabla^{\,K^{*}}\dr K^{\,*},\qquad\mbox{and}\qquad 
h^{\,U}=\nabla^{\,U^{*}}\dr U^{\,*},
\label{definition-dual-connections-K-U}
\eeq
are referred to as the connection associated with $K^{\,*}$, and referred to 
as that associated  with $U^{\,*}$, respectively.
\end{Def}
Let $(\cN,g)$ be a Riemannian manifold, and $\nabla$ a connection such that 
there exists a coordinate system so that connection components vanish.  
Such coordinate system is referred to as a $\nabla$-affine coordinate system.
If there exists a function $\psi$ on $\cN$ such that $g=\nabla \dr\psi$,
then $(\cN,\nabla,g)$ is referred to as a Hessian manifold. 

By definition, $\nabla^{\,K}$ and  $\nabla^{\,U}$ are flat connections, where 
$\nabla^{K}$-affine coordinates are $\{p_{\,a}\}$, and $\nabla^{\,U}$-affine
ones are $\{q^{\,a}\}$. 
Then the triplets 
$(\cM_{\,K},\nabla^{\,K}, h^{\,K})$  and $(\cM_{\,U},\nabla^{\,U}, h^{\,U})$ 
are Hessian manifolds. Similarly, 
$\nabla^{\,K^{*}}$ and  $\nabla^{\,U^{*}}$ are flat connections, where 
$\nabla^{\,K^{*}}$-affine coordinates are $\{p_{\,*}^{\,a}\}$, and 
$\nabla^{\,U^{*}}$-affine ones are $\{q_{\,a}^{\,*}\}$. 
Then the triplets 
$(\cM_{\,K},\nabla^{\,K^{*}}, h^{\,K})$  and $(\cM_{\,U},\nabla^{\,U^{*}}, h^{\,U})$ 
are Hessian manifolds.

There is some overlap between information geometry and 
Hessian geometry, and cubic forms are 
defined in information geometry. Such cubic forms also appear in rewriting 
Hamilton's equations. 
\begin{Def}
{\bf (Cubic form).}\ The following $(0,3)$-tensor fields 
$$
C^{\,K}=\nabla^{\,K}h^{\,K},\qquad \mbox{and}\qquad 
C^{\,U}=\nabla^{\,U}h^{\,U},
$$
are referred to as the cubic form associated with $K$ and referred to as 
that with $U$, respectively. Similarly, 
$$
C^{\,K^{*}}=\nabla^{\,K^{*}}h^{\,K},\qquad \mbox{and}\qquad 
C^{\,U^{*}}=\nabla^{\,U^{*}}h^{\,U},
$$
are referred to as the cubic form associated with $K^{\,*}$ and referred to as 
that with $U^{\,*}$, respectively.
\end{Def}
Note that cubic forms are not $3$-forms. 
The components of cubic form are given as follows. 
\begin{Lemma}
\label{fact-components-cubic-forms}
In terms of $\nabla^{\,K}$-affine coordinates $\{p_{\,a}\}$ and 
$\nabla^{\,U}$-affine coordinates $\{q^{\,a}\}$, 
the components of the cubic forms 
$$
C^{\,K}=C_{\,K}^{\,abc}\,\dr p_{\,a}\otimes \dr p_{\,b}\otimes \dr p_{\,c},
$$
and 
$$
C^{\,U}=C_{\, abc}^{\,U}\,\dr q^{\,a}\otimes \dr q^{\,b}\otimes \dr q^{\,c},
$$
are written as 
$$
C_{\,K}^{\,abc}
=\frac{\partial^3\, K}{\partial p_{\,a}\partial p_{\,b}\partial p_{\,c}}
,\qquad\mbox{and}\qquad 
C_{\,abc}^{\,U }
=\frac{\partial^3\, U}{\partial q^{\,a}\partial q^{\,b}\partial q^{\,c}}.
$$
Similarly, 
in terms of $\nabla^{\,K^{*}}$-affine coordinates $\{p_{\,*}^{\,a}\}$ and 
$\nabla^{\,U^{*}}$-affine 
coordinates $\{q_{\,a}^{\,*}\}$, the components of the cubic forms 
$$
C^{\,K^{*}}
=C_{\,abc}^{\,K^{*}}\,
\dr p_{\,*}^{\,a}\otimes \dr p_{\,*}^{\,b}\otimes \dr p_{\,*}^{\,c},
$$
and 
$$
C^{\,U^{*}}
=C_{\, U^{*}}^{\,abc}\,\dr q_{\,a}^{\,*}\otimes \dr q_{\,b}^{\,*}\otimes \dr q_{\,c}^{\,*},
$$
are written as 
$$
C_{\,abc}^{\,K^{*}}
=\frac{\partial^3\, K^{\,*}}{\partial p_{\,*}^{\,a}\,\partial p_{\,*}^{\,b}\,\partial 
p_{\,*}^{\,c}},\qquad\mbox{and}\qquad 
C_{\,U^{*}}^{\, abc}
=\frac{\partial^3\, U^{\,*}}{\partial q_{\,a}^{\,*}\,\partial q_{\,b}^{\,*}\,\partial q_{\,c}^{\,*}}.
$$
\end{Lemma}
\begin{Proof}
Let $X,Y,Z$ be vector fields whose basis is $\{\partial/\partial p_{\,a}\}$. 
Then one has
$$
C^{\,K}(X,Y,Z)
=\left(\,\nabla_{\,X}^{K}h^{\,K}\,\right)(Y,Z)
=X(\,h^{\,K}(Y,Z)\,)
-h^{\,K}(\nabla_{\,X}^{K}Y,Z)-h^{\,K}(Y,\nabla_{\,X}^{K}Z).
$$
Since $\{p_{\,a}\}$ is a set of $\nabla^{K}$-affine coordinates, one has 
$\nabla_{\,X}^{K}Y=0$ and $\nabla_{\,X}^{K}Z=0$. Combining these and
\fr{metric-h-components}, one arrives at  
$$
C_{\,K}^{\, abc}
=C^{\,K}\left(\,
\frac{\partial}{\partial p_{\,a}},\frac{\partial}{\partial p_{\,b}},
\frac{\partial}{\partial p_{\,c}}\,\right)
=\frac{\partial\, h_{\,K}^{\,bc}}{\partial p_{\,a}}
=\frac{\partial^3\, K}{\partial p_{\,a}\partial p_{\,b}\partial p_{\,c}}.
$$
Similarly one can calculate the components  
$C_{\,abc}^{\,U}, C_{\,abc}^{\,K^{*}}$ and $C_{\,U^{*}}^{\, abc}$.
\qed
\end{Proof}
As shown below, 
the components of the cubic forms $C^{\,K}$ and $C^{\,U}$ are related to 
the connection components of the Levi-Civita connections associated with 
$h^{\,K}$ and $h^{\,U}$. 
\begin{Lemma}
\label{fact-Levi-Civita-cubic-form-components}
Let $\nabla^{\,K\,(0)}$ and $\nabla^{\,U\,(0)}$ be 
the Levi-Civita connections associated with $h^{\,K}$ and $h^{\,U}$, 
$\{ \Gamma_{K(0)\, c}^{\,ab }\}$ connection coefficients for 
$\nabla^{\,K\,(0)}$ such that 
$\nabla_{\partial^{\,a}}^{\,K\,(0)}\partial^{\,b}=\Gamma_{\,K(0)\,c}^{\, ab}\partial^{\,c}$, 
$(\partial^{\,a}:=\partial/\partial p_{\,a})$,  
$\{ \Gamma_{\,ab}^{\,U(0) \, c}\}$ connection coefficients for 
$\nabla^{\,U\,(0)}$ such that 
$\nabla_{\partial_a}^{\,U\,(0)}\partial_{b}=\Gamma_{\,ab}^{\,U(0) \, c}\partial_{c}$, 
$(\partial_{a}:=\partial/\partial q^{\,a})$. Then 
$\Gamma_{\,K(0)}^{\,abc}:=h_{\,K}^{\,cj}\,\Gamma_{\,K(0)\,j}^{\,ab}$ and 
$\Gamma_{\,abc}^{\,U(0)}:=h_{\,cj}^{\,U}\,\Gamma_{\,ab}^{\,U(0)\,j}$ are given by 
$$
\Gamma_{\,K(0)}^{\,abc}
=\frac{1}{2}
\frac{\partial^3\,K}{\partial p_{\,a}\partial p_{\,b}\partial p_{\,c} }
=\frac{1}{2}C_{\,K}^{\,abc},
\qquad\mbox{and}\qquad 
\Gamma_{\,abc}^{\,U(0)}
=\frac{1}{2}
\frac{\partial^3\,U}{\partial q^{\,a}\partial q^{\,b}\partial q^{\,c} }
=\frac{1}{2}C_{\,abc}^{\,U}.
$$
Also, let 
$\{ \Gamma_{\,ab}^{\, K^{*}(0)\, c}\}$ be connection coefficients for 
$\nabla^{\,K\,(0)}$ such that 
$\nabla_{\partial_{\,a}}^{\,K^{*}\,(0)}\partial_{\,b}
=\Gamma_{\,ab}^{\,K^{*}(0)\,c }\partial_{\,c}$, 
$(\partial_{\,a}:=\partial/\partial p_{\,*}^{\,a})$,  
$\{ \Gamma_{\,U^{*}(0)\, c}^{\,ab }\}$ connection coefficients for 
$\nabla^{\,U\,(0)}$ such that 
$\nabla_{\partial^{\,a}}^{\,U^{*}\,(0)}\partial^{\,b}
=\Gamma_{\,U^{*}(0) \, c}^{\,ab}\partial^{\,c}$, 
$(\partial^{\,a}:=\partial/\partial q_{\,a}^{\,*})$. Then 
$\Gamma_{\,abc}^{\,K^{*}(0)}:=h_{\,cj}^{\,K}\,\Gamma_{\,ab}^{\,K^{*}(0)\,j}$ and 
$\Gamma_{\,U^{*}(0)}^{\,abc}:=h_{\,U}^{\,cj}\,\Gamma_{\,U^{*}(0)\,j}^{\,ab}$ are given by 
$$
\Gamma_{\,abc}^{\,K^{*}(0)}
=\frac{1}{2}
\frac{\partial^3\,K^{\,*}}{\partial p_{\,*}^{\,a}\partial\, p_{\,*}^{\,b}\,
\partial p_{\,*}^{\,c} }
=\frac{1}{2}C_{\,abc}^{\,K^{*}},
\qquad\mbox{and}\qquad 
\Gamma_{\,U^{*}(0)}^{\,abc}
=\frac{1}{2}
\frac{\partial^3\,U^{\,*}}{\partial q_{\,a}^{\,*}\,\partial q_{\,b}^{\,*}\partial q_{\,c}^{\,*} }
=\frac{1}{2}C_{\,U^{*}}^{\,abc}.
$$ 
\end{Lemma}
\begin{Proof}
A proof for 
$\Gamma_{\,abc}^{\,U(0)}$ is given as follows.
Substituting $h_{\,ab}^{\,U}=\partial^2\, U/\partial p^{\,a}\partial p^{\,b}$ into 
$$
\Gamma_{\,abc}^{\,U(0)}
=h_{\,cj}^{\,U}\Gamma_{\,ab}^{\,U(0)\,j}
=\frac{1}{2}\left(
\frac{\partial h_{\,cb}^{\,U}}{\partial q^{\,a}}
+\frac{\partial h_{\,ac}^{\,U}}{\partial q^{\,b}}
-\frac{\partial h_{\,ab}^{\,U}}{\partial q^{\,c}}
\right),
$$
one has 
$$
\Gamma_{\,abc}^{\,U(0)}
=\frac{1}{2}
\frac{\partial^3\,U}{\partial q^{\,a}\partial q^{\,b}\partial q^{\,c} }.
$$
Combining this with lemma\,\ref{fact-components-cubic-forms}, 
one has $\Gamma_{\,abc}^{\,U(0)}=C_{\,abc}^{\,U}/2$.   
Similarly proofs for 
$\Gamma_{\,K(0)}^{\,abc},\Gamma_{\,abc}^{\,K^{*}(0)}$ and $\Gamma_{\,U^{*}(0)}^{\,abc}$ 
can be given.
\qed
\end{Proof}

The canonical equations of motion \fr{canonical-equations-general}
can then be written as 
\beq
\frac{\dr}{\dr t}\left(\,
\frac{\partial\, U^{\,*}}{\partial q_{\,a}^{\,*}}
\,\right)
=p_{\,*}^{\,a},\qquad\mbox{and}\qquad 
\frac{\dr}{\dr t}\left(\,
\frac{\partial K^{\,*}}{\partial p_{\,*}^{\,a}}
\,\right)
=-\,q_{\,a}^{\,*},\qquad a\in\{1,\ldots,n\}.
\label{canonical-equations-general-first-order}
\eeq
It should be noted that the force term $-\partial\,U/\partial q_{\,a}$ in 
the original coordinate system is 
linear $-q_{\,a}^{\,*}$ in the dual coordinate system.
This linearization scheme for force term 
may be seen as an extension or a variant of 
Toda's dual transform\,\cite{Toda1967} 
( see section\,\ref{section-toda-dual} ). 

From the viewpoint above, 
one arrives at the following set of transformed equations,  
and this is summarized as the main theorem in this paper. 
\begin{Thm}
\label{fact-generalized-Toda-dual-transformed-equations}
{\bf (Generalized Toda's dual transformed equations).}\  
Consider the natural Hamiltonian system 
\fr{canonical-equations-general}. If $\{\,h_{\,K}^{\,ab}\,\}$ are constant, then 
the canonical equations are written as
\beq
\frac{\dr}{\dr t}
\left(\,\left.
\frac{\partial U^{\,*}}{\partial q_{\,a}^{\,*}}
\right|_{q_{\,a}^{\,*}=-\dot{p}_{\,a}}\,\right)
=\,h_{\,K}^{\,ab}\,p_{\,b}+h_{\,K}^{\,a\, (0)},\qquad 
a\in\{1,\ldots,n\},
\label{fact-generalized-Toda-lattice}
\eeq
where $\dot{p}_{\,a}:=\dr p_{\,a}/\dr t$ and $\{h_{\,K}^{\,a\,(0)}\}$ are constant.
\end{Thm}
\begin{Proof}
One has second order equations of motion 
in the transformed coordinates as follows. 
It follows from \fr{canonical-equations-general}  that 
$$
\frac{\dr^2 q^{\,a}}{\dr t^{\,2}}
=-\frac{\partial^2 K}{\partial p_{\,a}\partial p_{\,b}}
\frac{\partial\,U}{\partial q^{\,b}},\qquad 
a\in\{1,\ldots,n\}
$$
from which 
\beq
\frac{\dr^2}{\dr t^2}\left(\frac{\partial\,U^{\,*}}{\partial q_{\,a}^{\,*}}
\right)
=-h_{\,K}^{\,ab}q_{\,b}^{\,*},\qquad 
a\in\{1,\ldots,n\}.
\label{canonical-equations-general-second-order} 
\eeq 
Substituting $\dot{p}_{\,a}=-\,q_{\,a}^{\,*}$ coming from the second equation of 
\fr{canonical-equations-general-first-order} into 
\fr{canonical-equations-general-second-order}, one has
$$
\frac{\dr^2}{\dr t^2}\left(\,\left.
\frac{\partial\,U^{\,*}}{\partial q_{\,a}^{\,*}}\right|_{q_{\,a}^{\,*}=-\,\dot{p}_{\,a}}
\,\right)
=h_{\,K}^{\,ab}\,\frac{\dr p_{\,b}}{\dr t}.
$$ 
Since $\{h_{\,K}^{\,ab}\}$ are constant, 
one can integrate the equations above with respect to 
$t$. These calculations yield \fr{fact-generalized-Toda-lattice}.
\qed
\end{Proof}
In this paper the set of equation   
\fr{canonical-equations-general-second-order} is referred to as the 
generalized Toda's dual transformed equations, and it will be shown 
how these transformed equations are related to the original Toda's equations in 
section\,\ref{section-toda-dual}.

One can integrate \fr{canonical-equations-general-second-order}
once more by changing variables. 
Choice of such new variables depends on the given system. However 
the following change of variables is a generalization for the case of 
the Toda lattice system.
First,  
the abbreviation
$$
U^{\,*\,a}(q^{\,*})
:=\frac{\partial U^{\,*}}{\partial q_{\,a}^{\,*}},\qquad 
a\in\{1,\ldots,n\},
$$
is introduced, then one introduces the variables 
$\tau^{\,U}=\{\tau_{\,a}^{\,U}\}$ 
that depend on $t$ as 
$$
p_{\,a}(\tau^{\,U})
=\delta_{\,ab}\frac{\dr }{\dr t}U^{\,*\,b}(\tau^{\,U}).
$$
It follows that 
$$
p_{\,a}(\tau^{\,U}(t))
=\delta_{\,ab}\,h_{\,U}^{\,bc}\,\frac{\dr\tau_{\,c}}{\dr t},\qquad\mbox{and}\qquad 
\frac{\dr}{\dr t}p_{\,a}(\tau(t)) 
=\delta_{\,ab}\,\frac{\dr^2\,U^{\,*\,b}}{\dr t^2}
=\delta_{\,ab}\,\left(\frac{\partial\, h_{\,U}^{\,bc}}{\partial \tau_{\,i}}
\frac{\dr\tau_{\,c}^{\,U}}{\dr t}\frac{\dr\tau_{\,i}^{\,U}}{\dr t}
+h_{\,U}^{\,bc}\,\frac{\dr^2\tau_{\,c}^{\,U}}{\dr t^2}\right),
$$
where $\delta_{\,ab}$ is  
the Kronecker delta giving unity for $a=b$, and zero 
otherwise. 
In these coordinates, one has from 
\fr{fact-generalized-Toda-lattice}
that 
$$
U^{\,*\,a}\left(-\,\delta_{\,ab}\frac{\dr^2\,U^{\,*\,b}}{\dr t^2}\right)
=h_{\,K}^{\,ab}\,\delta_{\,bc}\,U^{\,*\,c}(\tau^{\,U})+h_{\,K}^{\,a\,(0)}\,t
+h_{\,K}^{\,a\,(0,0)},
$$
where $\{h_{\,K}^{\,a\,(0,0)}\}$ are constant.
For the case of the Toda lattice system, $U^{\,*}$ is a logarithm function, 
and the functions $\tau^{\,U}$ is the so-called $\tau$-functions 
( see section\,\ref{section-toda-dual} ).

The following states how the canonical equations of motion are written 
in terms of Hessian-information geometry.
\begin{Thm}
\label{fact-canonical-equation-Hessian-language}
{\bf (Canonical equations of motion written in terms of Hessian geometry).}\   
The canonical equations of motion in terms of the dual coordinates 
\fr{canonical-equations-general-second-order} 
can be written in the forms 
\beq
C_{\,U^{*}}^{\,abc}\,\dot{q}_{\,b}^{\,*}\dot{q}_{\,c}^{\,*}
+h_{\,U}^{\,ab}\ddot{q}_{\,b}^{\,*}
=-h_{\,K}^{\,ab}q_{\,b}^{\,*},\qquad 
a\in\{1,\ldots,n\},
\label{canonical-equations-general-second-order-cubic-form}
\eeq
or equivalently
$$
2\Gamma_{\,U^{*}(0)}^{\,abc}\,\dot{q}_{\,b}^{\,*}\dot{q}_{\,c}^{\,*}
+h_{\,U}^{\,ab}\ddot{q}_{\,b}^{\,*}
=-h_{\,K}^{\,ab}q_{\,b}^{\,*},\qquad 
a\in\{1,\ldots,n\},
$$
where $\dot{q}_{\,a}^{\,*}:=\dr q_{\,a}^{\,*}/\dr t$ and 
$\ddot{q}_{\,a}^{\,*}:=\dr^2 q_{\,a}^{\,*}/\dr t^2$.
\end{Thm}
\begin{Proof}
First, \fr{canonical-equations-general-second-order} 
can be written as 
$$
\frac{\partial^3\, U^{\,*}}{\partial q_{\,a}^{\,*}\,\partial q_{\,b}^{\,*}\,\partial q_{\,c}^{\,*}}
\dot{q}_{\,c}^{\,*}\dot{q}_{\,b}^{\,*}
+\frac{\partial^2\, U^{\,*}}{\partial q_{\,a}^{\,*}\,\partial q_{\,b}^{\,*}}\ddot{q}_{\,b}^{\,*}
=-h_{\,K}^{\,ab}q_{\,b}^{\,*},\qquad 
a\in\{1,\ldots,n\}.
$$
Then 
substituting the explicit forms of $C_{\,U^{*}}^{\,abc}$, 
$h_{\,U}^{\,ab},h_{\,K}^{\,ab}$, and $\Gamma_{\,U^{*}(0)}^{\,abc}$ obtained in 
lemmas\,    
\ref{fact-components-cubic-forms} and 
\ref{fact-Levi-Civita-cubic-form-components} 
into the equations above, 
one completes the proof.
\qed
\end{Proof}
In addition to this theorem, one can see how the bases 
$\{\partial/\partial q^{\,a}\}$,$\{\partial/\partial q_{\,a}^{\,*}\}$,  
$\{\partial/\partial p_{\,a}\}$, and $\{\partial/\partial p_{\,*}^{\,a}\}$
are oriented. 
From discussions in information geometry\,\cite{Goto2016,BN2016}, 
they are 
$$
h^{\,K}\left(
\frac{\partial}{\partial p_{\,a}},
\frac{\partial}{\partial p_{\,*}^{\,b}}\right)
=\delta_{\,b}^{\,a},\qquad\mbox{and}\qquad 
h^{\,U}\left(
\frac{\partial}{\partial q^{\,a}},
\frac{\partial}{\partial q_{\,b}^{\,*}}\right)
=\delta_{\,a}^{\,b},
$$
In addition, the solution to the canonical equations of motion satisfy 
$q_{\,a}^{*}=-\,\dot{p}_{\,a}$. This yields that 
the vector $\partial/\partial q^{\,a}$ is anti-parallel to 
$\partial/\partial \dot{p}_{\,a}$.

Extending $\Gamma_{\,U^{*} (0)}^{\,abc}$, one can have the 
one-parameter family of connection coefficients as 
\beq
\Gamma_{\,U^{*} (\alpha)}^{\,abc}
:=\frac{1-\alpha}{2}\frac{\partial^3\,U^{\,*}}{\partial q_{\,a}^{\,*}\,
\partial q_{\,b}^{\,*}\,\partial q_{\,c}^{\,*}},\qquad \alpha\in\mbbR.
\label{pseudo-alpha-connection-components}
\eeq
It follows from  \fr{pseudo-alpha-connection-components} that  
$\Gamma_{\,U^{*} (\alpha)}^{\,abc}$ 
and $\Gamma_{\,U^{*} (-\alpha)}^{\,abc}$ satisfy  
$$
\Gamma_{\,U^{*} (\alpha)}^{\,abc}
+\Gamma_{\,U^{*} (-\alpha)}^{\,abc}
=\frac{\partial}{\partial q_{\,a}^{\,*}}h_{\,U}^{\,bc}.
$$
In the context of information geometry, 
the pair of connection coefficients 
$\Gamma_{\,U^{*} (\alpha)}^{\,abc}$ and $\Gamma_{\,U^{*} (-\alpha)}^{\,abc}$
are referred to as the components of dual connections with respect to 
$h_{\,U}$\,\cite{AN}. 
It should be noted that the $\alpha$-connection plays a role in information 
geometry. As shown below, 
this family of connections can also appear in this geometric formulation of 
classical Hamiltonian systems.

\begin{Proposition}
The canonical equations of motion are written in terms of the 
$\alpha$-connection with $\alpha=-1$ as  
\beq
\frac{\dr^2 q_{\,i}^{\,*}}{\dr t^{\,2}}
+h_{\,ij}^{\,U}\Gamma_{\,U^{*}(-1)}^{\,jbc}\,\frac{\dr q_{\,b}^{\,*}}{\dr t}
\frac{\dr q_{\,c}^{\,*}}{\dr t}
=-h_{\,ij}^{\,U}h_{\,K}^{\,jb} q_{\,b}^{\,*},\qquad 
i\in\{1,\ldots,n\}.
\label{canonical-equations-general-second-minus-1-form}
\eeq

\end{Proposition}
\begin{Proof}
The canonical equations of motion 
\fr{canonical-equations-general-second-order} 
can be written with $\Gamma_{\,U^{*}(-1)}^{\,abc}$ as follows. 
Combining lemma\,\ref{fact-Levi-Civita-cubic-form-components} and 
\fr{pseudo-alpha-connection-components}, one has 
\beq
C_{\,U^{*}}^{\,abc}
=2\Gamma_{\,U^{*}(0)}^{\,abc}
=\Gamma_{\,U^{*}(-1)}^{\,abc}.
\label{relation-Cubic-Gamma-0-1}
\eeq 
Substituting \fr{relation-Cubic-Gamma-0-1}
into \fr{canonical-equations-general-second-order-cubic-form}, 
one has 
\fr{canonical-equations-general-second-minus-1-form}. 
\qed
\end{Proof}
One can also write 
\fr{canonical-equations-general-second-minus-1-form} 
as   
\beq
\frac{\dr^2 q_{\,i}^{\,*}}{\dr t^{\,2}}
+h_{\,ij}^{\,U}h_{\,K}^{\,jb} q_{\,b}^{\,*}
=-h_{\,ij}^{\,U}\Gamma_{\,U^{*}(-1)}^{\,jbc}\,\frac{\dr q_{\,b}^{\,*}}{\dr t}
\frac{\dr q_{\,c}^{\,*}}{\dr t},\qquad 
i\in\{1,\ldots,n\}.
\label{canonical-equations-general-second-minus-1-form-2}
\eeq
This form of the equations is similar to a form of 
a perturbed harmonic oscillator if  $\Gamma_{\,U^{*}(-1)}^{\,jbc}$ are small enough.
A variety of applications of perturbed harmonic oscillators 
are found in physics, and a solution to the unperturbed system is 
the basis of discussion in general.   
Thus the case where $\Gamma_{\,U^{*}(-1)}^{\,jbc}$ vanish is of interest.   

As shown below, the following 
duality holds in the case where $U$ is quadratic.  
\begin{Proposition}
Consider the case where $U$ is quadratic.
Then a solution to the canonical equations of motion is also a solution to the 
equations obtained by replacing the $(-1)$-connection with $(1)$-connection
in \fr{canonical-equations-general-second-minus-1-form}. 
\end{Proposition}
\begin{Proof}
  Throughout this proof, 
the set of variables $\{\wh{q}_{\,a}^{\,*}\}$ 
is introduced in order to emphasize that a quadratic potential is focused, and 
$\{\wh{q}_{\,a}^{\,*}\}$ is distinguished from   $\{q_{\,a}^{\,*}\}$. Similarly 
$\{\wh{q}^{\,a}\}$ is introduced. 
Since $U$ is quadratic, one has from \fr{metric-h-components} that  
$$
U(\wh{q})
=\frac{1}{2}h_{\,ab}^{\,U}(0)\,\wh{q}^{\,a}\wh{q}^{\,b}.
$$
Then the Legendre transform of $U$ is calculated 
from \fr{Legendre-transform-K-U} as 
$$
U^{\,*}(\wh{q}^{\,*})
=\sup_{q}\left[\,\wh{q}^{\,a}\,\wh{q}_{\,a}^{\,*}-U(\,\wh{q}\,)\,\right]
=\left[\,\wh{q}^{\,a}\,\wh{q}_{\,a}^{\,*}-U(\,\wh{q}\,)\,
\right]_{\,\wh{q}^a=h_{\,U}^{\,ai}(0)\,\wh{q}_{\,i}^{\,*}}
=\frac{1}{2}h_{\,U}^{\,ab}(0)\,\wh{q}_{\,a}^{\,*}\wh{q}_{\,b}^{\,*}.
$$
For $U^{\,*}$, it follows from 
\fr{pseudo-alpha-connection-components} that 
$\Gamma_{\,U^{*}(-1)}^{\,abc}\equiv 0$. 
With this, $h_{\,ij}^{\,U}=h_{\,ij}^{\,U}(0)$, and 
\fr{canonical-equations-general-second-minus-1-form-2}, one has 
\beq
\frac{\dr^2\, \wh{q}_{\,i}^{\,*}}{\dr t^{\,2}}
+h_{\,ij}^{\,U}(0)h_{\,K}^{\,jb}\, \wh{q}_{\,b}^{\,*}
=0,\qquad 
i\in\{1,\ldots,n\}. 
\label{canonical-equations-general-second-minus-1-form-homogeneous}
\eeq

Consider 
the equations obtained by 
replacing $\Gamma_{\,U^{*}(-1)}^{\,abc}$ with $\Gamma_{\,U^{*}(1)}^{\,abc}$ in 
\fr{canonical-equations-general-second-minus-1-form-2} :  
\beq
\frac{\dr^2 \wh{q}_{\,i}^{\,*(1)}}{\dr t^{\,2}}
+h_{\,ij}^{\,U}h_{\,K}^{\,jb}\wh{q}_{\,b}^{\,*(1)}
=-h_{\,ij}^{\,U}\Gamma_{\,U^{*}(1)}^{\,jbc}\,
\frac{\dr \wh{q}_{\,b}^{\,*(1)}}{\dr t}
\frac{\dr \wh{q}_{\,c}^{\,*(1)}}{\dr t},\qquad 
i\in\{1,\ldots,n\},
\label{canonical-equations-general-second-plus-1-form}
\eeq
where the set of variables $\{\wh{q}_{\,a}^{\,*\,(1)}\}$ has been introduced 
to emphasize that the $\alpha$-connection with $\alpha=1$ 
has been focused.  
It follows from \fr{pseudo-alpha-connection-components} that 
$\Gamma_{\,U^{*}(1)}^{\,jbc}\equiv 0$.   
 Then with  $h_{\,ij}^{\,U}=h_{\,ij}^{\,U}(0)$ 
the equations 
\fr{canonical-equations-general-second-plus-1-form} 
 reduce to 
\beq
\frac{\dr^2 \wh{q}_{\,i}^{\,*(1)}}{\dr t^{\,2}}
+h_{\,ij}^{\,U}(0)h_{\,K}^{\,jb}\,\wh{q}_{\,b}^{\,*(1)}
=0,\qquad 
i\in\{1,\ldots,n\}.
\label{canonical-equations-general-second-plus-1-form+1}
\eeq
Thus, a solution of \fr{canonical-equations-general-second-plus-1-form+1}
is that of 
\fr{canonical-equations-general-second-minus-1-form-homogeneous}.
\qed
\end{Proof}
\begin{Remark}
If $K$ is also quadratic, then  
\fr{canonical-equations-general-second-minus-1-form-homogeneous} 
is a set of linear equations.
\end{Remark}


Before closing this subsection, the present geometric formulation 
is compared with the one proposed by Teruel\,\cite{Teruel2013}.
 
Define the function $J$ that was originally introduced in 
\cite{Teruel2013}  as  the total Legendre transform of $H$   
$$
J(\dot{q},\dot{p})
:=\inf_{q,p}\left[\,
\dot{p}_{\,a}\,q^{\,a}-\dot{q}^{\,a}p_{\,a}+H(q,p)
\,\right],
$$
from which 
$$
J(\dot{q},\dot{p})
=-\sup_{p,q}\left[\,
\dot{q}^{\,a}p_{\,a}-\dot{p}_{\,a}\,q^{\,a}-U(q)-K(p) 
\,\right].
$$
In the dual coordinates introduced in 
definition\,\ref{definition-dual-coordinates-natural-Hamiltonian}
$$
p_{\,*}^{\,a}
=\frac{\partial K}{\partial p_{\,a}}
=\dot{q}^{\,a},\qquad\mbox{and}\qquad 
q_{\,a}^{\,*}
=\frac{\partial U}{\partial q^{\,a}}
=-\,\dot{p}_{\,a},
$$
where \fr{canonical-equations-general} has been used,  
one can express $J$ as 
$$
J(p_{\,*},-\,q^{\,*})
=-\sup_{q,p}\left[\,
p_{\,a}^{\,*}\,p_{\,a}+q_{\,a}^{\,*}\,q^{\,a}-U(q)-K(p) 
\,\right].
$$
With this and \fr{Legendre-transform-K-U}, one has 
\beq
J(p_{\,*},-\,q^{\,*})
=-\,\left[\,K^{\,*}(p_{\,*})+U^{\,*}(q^{\,*})\,\right].
\label{J-by-K-U}
\eeq 
This states how $J$ in \cite{Teruel2013} is related to 
the present formulation.  
Hamilton's equations can also be written with $J$ as  follows.
Applying 
\fr{coordinates-by-dual-coordinates} to 
differentiation of \fr{J-by-K-U}, one has
$$
\frac{\partial J}{\partial\, p_{\,*}^{\,a}}
=-\,\frac{\partial K^{\,*}}{\partial\, p_{\,*}^{\,a}}
=-\,p_{\,a},\qquad\mbox{and}\qquad 
\frac{\partial J}{\partial\, q_{\,a}^{\,*}}
=-\,\frac{\partial U^{\,*}}{\partial\, q_{\,a}^{\,*}}
=-\,q^{\,a},
$$
from which 
$$
\frac{\dr}{\dr t}\left(\,\frac{\partial J}{\partial\, p_{\,*}^{\,a}}
\right)
=-\,\dot{p}_{\,a}
=q_{\,a}^{\,*},\qquad \mbox{and}\qquad 
\frac{\dr}{\dr t}\left(\,\frac{\partial J}{\partial\, q_{\,a}^{\,*}}
\right)
=-\,\dot{q}^{\,a}
=-\,p_{\,*}^{\,a},\qquad a\in\{1,\ldots,N\}.
$$
These derived equations correspond to 
\fr{canonical-equations-general-first-order}.

\subsection{Relation  to Toda's original dual transform}
\label{section-toda-dual}
In this subsection it is shown how Toda's original dual transform is related to 
the geometric formulation introduced 
in section\,\ref{section-non-vanishing-potential-systems}.
To this end, 
Toda's original idea of the dual transform and the dual lattice system 
are briefly 
reviewed below first. 

Toda considered in \cite{Toda1967} Hamilton's equations of motion
for the $a$-th particle ($a\in \{0,\ldots,N\}$) in a chain 
\beq
m\,\frac{\dr^2 u^{\,a}}{\dr t^2}
=-\phi'(u^{\,a}-u^{\,a-1})+\phi'(u^{\,a+1}-u^{\,a}),
\label{Toda-starting-point-lattice}
\eeq 
where $m>0$ stands for the mass of the particles, $u_{\,a}$
the position of the $a$-th particle, 
$\phi$ the interaction 
potential energy function depending on the distance between adjacent 
particles, and 
${}^{\prime}$ its derivative with respect to the argument. 
In this paper the system \fr{Toda-starting-point-lattice}
is referred to as the original lattice system. 
Then the potential 
energy function $U$ is the sum of $\phi$. 
Introducing the variables 
$$
q_{\,a}
:=u^{\,a+1}-u^{\,a},\qquad\mbox{and}\qquad
f_{\,a}
:=-\,\phi'(q_{\,a}),\qquad a\in\{1,\ldots,N\}
$$
one can derive 
$$
\ddot{q}^{\,a}
=\frac{-1}{m}\left(\,f_{\,a+1}+f_{a-1}-2\,f_{\,a}\,\right),
$$
or equivalently   
with $p_{\,a}$ satisfying $\dot{p}_{\,a}=f_{\,a}$, $(a\in\{1,\ldots,N\})$,  
\beq
\dot{q}^{\,a}
=\frac{-1}{m}\left(\,p_{\,a+1}+p_{a-1}-2\,p_{\,a}\,\right),\qquad
\mbox{and}\qquad 
\dot{p}_{\,a}
=-\,\frac{\dr\,\phi(q^{\,a})}{\dr q^{\,a}}.
\label{Toda-dual-lattice-0}
\eeq
If one finds a function $\chi$ 
by solving the second equation of 
\fr{Toda-dual-lattice-0} for $q^{\,a}$ such that 
\beq
q^{\,a}
=-\,\frac{\chi(\,\dot{p}_{\,a}\,)}{m},
\label{Toda-chi-definition-general}
\eeq
for all $a\in\{\,0,\ldots,N\, \}$, then one has
\beq
\frac{\dr}{\dr t}\chi(\dot{p}_{\,a})
=p_{\,a+1}+p_{\,a-1}-2\,p_{\,a}.
\label{Toda-dual-lattice-general}
\eeq

Toda proposed the nonlinear form of $\phi$,  
\beq
\phi(q)
=\frac{A}{B}\,\e^{-B q}+A\, q,
\label{Toda-potential-AB}
\eeq
with $A$ and $B$ being positive constant. 
This interaction potential energy function $\phi$  
is referred to as the Toda potential function.  
In this case the function $\chi$ is obtained as 
$$
\chi(\,\dot{p}_{\,a}\,)
=\frac{m}{B}\ln\left(1+\frac{\dot{p}_{\,a}}{A}\right),
$$ 
and one has the Toda lattice system :  
\beq
m\,\frac{\dr}{\dr t}\left[\,\frac{1}{B}\ln\left(1+\frac{\dot{p}_{\,a}}{A}\right)
\,\right]
=p_{\,a+1}+p_{\,a-1}-2\,p_{\,a}.
\label{Toda-lattice}
\eeq
The lattice system \fr{Toda-dual-lattice-general} is referred to as 
the dual lattice 
system with respect to the original lattice system 
\fr{Toda-starting-point-lattice}.
Note that the meaning of dual in Toda's original theory was  
not directly related to  
that in Hessian-information geometry. 

In his derivation of the Toda lattice system 
reviewed above, one observes the following. 
\begin{itemize}
\item
The interaction potential energy function $\phi$ is strictly convex, since 
$$
\frac{\dr^2\phi}{\dr q^{\,2}}
=AB\,\e^{\,-Bq}>0.
$$
\item
The existence of the function $\chi$ 
in \fr{Toda-chi-definition-general} 
is a key to find the second order equations of the dual lattice system. 
\end{itemize}
These observations lead to the  following  
\begin{itemize}
\item
The potential energy function  is strictly convex, since the sum of strictly 
convex functions is also a strictly convex function.
\item
If an explicit form of the 
Legendre transform of the potential energy function is found, 
then this is a key to find the dual lattice system.
\end{itemize}

The following is the main theorem in this subsection and it states 
how Toda's dual lattice system can be written in terms of 
the Legendre transform 
of the interaction potential energy function. 
\begin{Thm}
\label{fact-Toda-fact-chi-phi-Legendre-transform}
{\bf (The Legendre transform of interaction potential for dual lattice).}\  
Consider \fr{Toda-dual-lattice-0}, where 
$\phi$ is strictly convex. Assume that  
the Legendre transform $\phi^{\,*}$ of $\phi$,   
is explicitly  written. 
Then the function $\chi$ in \fr{Toda-dual-lattice-general} is written 
in terms of $\dot{p}_{\,a}$ and $\phi^{\,*}$ as 
\beq
\chi(\dot{p}_{\,a})
=-\,m\,
\left.
\frac{\dr\phi^{\,*}}{\dr q_{\,a}^{\,*}}
\right|_{\,q_{\,a}^{\,*}=-\dot{p}_{\,a}},\qquad a\in\{1,\ldots,N\}.
\label{Toda-fact-chi-phi-Legendre-transform}
\eeq
\end{Thm}
\begin{Proof}
Since $\phi$ is strictly convex, one can define the dual 
coordinates 
$\{q_{\,a}^{\,*}\}$, and they are related to $\{\dot{p}_{\,a}\}$ such that   
\beq
q_{\,a}^{\,*}
=\frac{\dr\,\phi(q^{\,a})}{\dr q^{\,a}}
=-\dot{p}_{\,a}.
\label{Toda-dual-position-general}
\eeq
Due to the assumption that the Legendre transform $\phi^{\,*}$ of $\phi$ 
can explicitly be written,  one can write 
\beq
q^{\,a}
=\frac{\dr\phi^{\,*}}{\dr q_{\,a}^{\,*}}.
\label{Toda-dual-position-via-dual-potential-general}
\eeq
The relation between 
$q^{\,a}$ and $\dot{p}_{\,a}$ is obtained by combining 
\fr{Toda-dual-position-general} and 
\fr{Toda-dual-position-via-dual-potential-general} as 
$$
q^{\,a}
=\left.\frac{\dr\phi^{\,*}}{\dr q_{\,a}^{\,*}}\right|_{q_{\,a}^{\,*}=-\dot{p}_{\,a}}.
$$ 
Substituting this into $\chi(\dot{p}_{\,a})=-\,m\,q^{\,a}(\dot{p}_{\,a})$ 
coming from 
\fr{Toda-chi-definition-general}, one obtains 
\fr{Toda-fact-chi-phi-Legendre-transform}. 
\qed
\end{Proof}
\begin{Remark}
With this theorem, the system \fr{Toda-dual-lattice-general} 
is written as 
\beq
-m\,\frac{\dr}{\dr t}\left(
\left.
\frac{\dr\phi^{\,*}}{\dr q_{\,a}^{\,*}}\right|_{\,q_{\,a}^{\,*}=-\dot{p}_{\,a}}
\right)
=p_{\,a+1}+p_{\,a-1}-2\,p_{\,a},
\label{Toda-lattice-generalized}
\eeq
which is a generalization of \fr{Toda-lattice}.
Also, the equation \fr{Toda-lattice-generalized}
are obtained from  
theorem\,\ref{fact-generalized-Toda-dual-transformed-equations}
with $U=\sum_{i}\phi(q^{\,i})$,  
$K=\sum_{i}(p_{\,i+1}-p_{\,i})^{\,2}/(2m)$, and $\{\,h_{\,K}^{\,a\,(0)}\,\}=0$.
\end{Remark}
From this theorem  it turns out 
that one significance of the proposed  
Hessian-information geometric 
formulation of 
Hamiltonian systems is to give how to systematically 
obtain an explicit form of 
Toda's dual transformed equations from a given lattice system. 
Thus, roughly speaking, dual in the sense of Toda is equivalent to 
that in the sense of Legendre.

In analyzing an integrable system, 
a set of $\tau$-functions may be focused, and 
these functions are used for constructing soliton  
solutions\,\cite{Nakamura1994,Miwa2000}. 
For the Toda lattice system \fr{Toda-lattice} with $A=B=m=1$, 
a set of $\tau$-functions $\{\tau_{\,a}\}$ is such that 
\beq
p_{\,a}
=\frac{\dr}{\dr t}\ln\tau_{\,a}.\qquad a\in\{1,\ldots,N\}
\label{Toda-p-tau}
\eeq
From \fr{Toda-p-tau} and \fr{Toda-dual-position-general}, 
one has the relation between $q_{\,a}^{\,*}$ and $\tau_{\,a}$ :
$$
q_{\,a}^{\,*}
=-\,\frac{\dr^2}{\dr t^{\,2}}\ln\tau_{\,a}.
$$
Then, the equations written in terms of a set of $\tau$-functions 
$\{\tau_{\,a}\}$ are obtained as follows. 
One can integrate \fr{Toda-dual-lattice-general} 
with respect to $t$ by introducing 
$\{\tau_{\,a}\}$ defined in \fr{Toda-p-tau} as 
$$
-m\,\left.\chi(\dot{p}_{\,a})\right|_{\dot{p}_{\,a}=\dot{p}_{\,a}(\tau_{\,a})}
=\ln\left(\frac{\tau_{\,a+1}\tau_{\,a-1}}{\tau_{\,a}^{\,2}}\right)+\const,
\qquad 
\mbox{where}\qquad 
\dot{p}_{\,a}(\tau_{\,a})
=\frac{\tau_{\,a}\ddot{\tau}_{\,a}-(\dot{\tau_{\,a}})^{\,2}}{\tau_{\,a}^{\,2}}.
$$
The left hand sides of the equations above are written as 
$$
-m\,\left.\chi(\dot{p}_{\,a})\right|_{\dot{p}_{\,a}=\dot{p}_{\,a}(\tau_{\,a})}
=
-m\,\left.\frac{\dr\phi^{\,*}}{\dr q_{\,*}^{\,a}}
\right|_{q_{\,*}^{\,a}=-\dot{p}_{\,a}(\tau_{\,a})}
=\frac{m}{B}\ln\left(1+\frac{\dot{p}_{\,a}(\tau_{\,a})}{A}\right).
$$
For the case where  $m=A=B=1$ and $\const=0$, 
one derives 
the set of equations for $\{\tau_{\,a}\}$ : 
$$
\ddot{\tau}_{\,a}\tau_{\,a}-\left(\dot{\tau}_{\,a}\right)^{\,2}
=\tau_{\,a+1}\tau_{\,a-1}-(\tau_{\,a})^{\,2}.
$$ 

To state the applicability of  Hessian-information geometry for 
\fr{Toda-dual-lattice-0}, one considers 
a system \fr{Toda-dual-lattice-0} with $\phi$ being strictly convex under 
some wider boundary conditions. 
The system is rewritten as 
the Hamiltonian system
$$
\dot{q}^{\,a}
=\frac{\partial K}{\partial p_{\,a}},\qquad 
\mbox{and}\qquad 
\dot{p}_{\,a}
=-\,\frac{\partial U}{\partial q^{\,a}},
$$
where
\beq
K(p)=\frac{1}{2m}\sum_{i}(\,p_{\,i+1}-p_{\,i}\,)^2,\qquad
\mbox{and}\qquad 
U(q)=\sum_{i}\phi(q^{\,i}),
\label{Toda-K-U-general}
\eeq 
under some boundary conditions.
Note that $U$ is strictly convex. Then it follows that   
\beq
h_{\,K}^{\,ab}
=\frac{\partial^2 K}{\partial p_{\,a}\partial p_{\,b}}
=-\,\frac{1}{m}(\,
\delta^{\,b\,a+1}+\delta^{\,b\,a-1}-2\, \delta^{\,b\, a} 
\,). 
\label{Toda-h-K-ab}
\eeq
Thus, to apply the general theory developed in 
section\,\ref{section-non-vanishing-potential-systems} to this system,  
one needs to verify that the condition 
$(\partial^2 K/\partial p_{\,a}\partial p_{\,b})\succ 0$ holds under  
boundary conditions.  For example if $p_{i+3}=p_{\,i}$ for any $i$, then 
$$
h_{\,K}^{\,ab}
=\frac{\partial^2 K}{\partial p_{\,a}\partial p_{\,b}}
=\frac{1}{m}
\left(
\begin{array}{ccc}
2&-1&0\\
-1&2&-1\\
0&-1&2
\end{array}
\right).
$$
The eigenvalues of the matrix $(h_{\,K}^{\,ab})$ 
are obtained as $(2\pm \sqrt{2})/m$ and $2/m$, and 
they are positive due to $m>0$. It is known that if all the eigenvalues of 
a matrix $M$ are positive, then $M$ is positive definite, $M\succ 0$. 
Applying this, one has that 
$(h_{\,K}^{\,ab})\succ 0$. From this, $K$ is strictly convex.  
Thus,   
Hessian-information geometric formulation can be applied to this 
periodic system.
 
Consider the case where the 
Hessian-information geometric formulation can be 
applied to a system whose Hamiltonian is $H=K+U$ with $K$ and $U$ 
given by \fr{Toda-K-U-general}. 
Also choose $\phi$ to be the Toda potential, \fr{Toda-K-U-general}.  
In this case the set of equations of motion 
\fr{canonical-equations-general-second-order} 
becomes \fr{Toda-lattice}. Then the quantities used in 
\fr{canonical-equations-general-second-order-cubic-form} 
are shown below.
The coordinates being dual to $\{q_{\,a}\}$ and the components $h_{\,ab}^{\,U}$ 
are  
$$
q_{\,a}^{\,*}
=\frac{\dr\, \phi}{\dr q^{\,a}}
=-A\e^{\,-B\,q^{\,a}}+A,\qquad\mbox{and}\qquad 
h_{\,ab}^{\,U}
=AB\,\e^{\,- B\,q_{\,a}}\,\delta_{\,ab} 
=B(A-q_{\,a}^{\,*})\,\delta_{\,ab},\quad\mbox{(no sum)}
$$
The Legendre transform of $\phi$ is 
$$
\phi^{\,*}(q_{\,a}^{\,*})
=\sup_{q}\left[\,q_{\,a}q_{\,*}^{\,a}-\phi(q)\,\right]
=\frac{A-q_{\,a}^{\,*}}{B}
\left[\,\ln\left(1-\frac{q_{\,a}^{\,*}}{A}\right)-1\,\right],
\qquad\mbox{(no sum)}
$$
from which 
$$
q^{\,a}
=\frac{\dr\,\phi^{\,*}}{\dr q_{\,*}^{\,a}}
=\frac{-1}{B}\ln\left(\,1-\frac{q_{\,a}^{\,*}}{A}\,\right),\qquad 
h_{\,U}^{\,ab}
=\frac{1}{B}\frac{1}{A-q_{\,a}^{\,*}}\,
\delta^{\,ab},\quad\mbox{(no sum)},
$$
and 
$$
C_{\,U^{\,*}}^{\,abc}
=\left\{
\begin{array}{cc}
\frac{1}{B(A-q_{\,a}^{\,*})^2}&\mbox{for $a=b=c$}\\
0&\mbox{otherwise}\\
\end{array}
\right..
$$
The components of the Riemannian metric tensor field $h_K$ have been 
calculated as \fr{Toda-h-K-ab}.

It has not been known how to systematically obtain 
the dual lattice systems from 
given lattice systems, and only a few examples 
 have been known, where explicit expressions of dual lattice systems are 
obtained.
They include linear and the Toda lattice systems. 
In what follows, some other examples of nonlinear lattice systems are 
shown,  where  
dual lattice systems are explicitly expressed. 

\begin{Example}
Consider 
a class of lattice systems of the form \fr{Toda-starting-point-lattice}
with the interaction potential energy function $\phi_{\,\beta}$ given by
$$
\phi_{\,\beta}(q)
=\frac{(\,q^{\,2}\,)^{\,\beta}}{2\beta},\qquad 2\beta>1, 
$$
where $\beta$ is fixed. Observe that $\phi_{\,\beta}(q)\in\mbbR$ for any 
$q\in\mbbR$.

The second order equations of the dual lattice system can be obtained by 
applying 
theorem\,\ref{fact-Toda-fact-chi-phi-Legendre-transform} to 
this system. To this end, one calculates the Legendre transform  
\beq
\phi_{\,\beta}^{\,*}(q^{\,*})
=\sup_{q}\left[\,q\,q^{\,*}-\phi_{\,\beta}(q)\,\right]
=\frac{(\,(\,q^{\,*}\,)^{\,2}\,)^{\,\beta^{\,*}}}{2\beta^{\,*}},
\label{Legendre-transform-power-function-with-beta}
\eeq
where $\beta^{\,*}$ satisfies 
$$ 
\frac{1}{2\beta}+\frac{1}{2\beta^{\,*}} 
=1.
$$
With \fr{Legendre-transform-power-function-with-beta} and 
\fr{Toda-lattice-generalized}, 
the explicit form of the dual lattice system is obtained as 
$$
-\,m\,\frac{\dr}{\dr t}
\left(\,(\,-\,\dot{p}_{\,a}\,)^{\,2}\,\right)^{\,\beta^{\,*}-1/2}
=p_{\,a+1}+p_{\,a-1}-2\,p_{\,a}.
$$
Note that the dual lattice system obtained above is linear 
($2\beta^{\,*}=2$)  
if the original lattice system is liner, $2\beta=2$.

The Hamiltonian written in terms of $\{\,q^{\,a}\,\}$ and $\{\,p_{\,a}\,\}$
is 
$$
H(p,q)=K(p)+U(q),\quad\mbox{where}\quad 
K(p)=\frac{1}{2m}\sum_{i}(p_{\,i+1}-p_{\,i})^{\,2},
\quad\mbox{and}\quad
U(q)=\sum_{i}\phi_{\,\beta}(\,q^{\,i}\,),\quad 2\beta>1.
$$
Consider the case where the Hessian-information geometric formulation can
be applied to this system. Then the quantities used in
 \fr{canonical-equations-general-second-order-cubic-form} 
are shown below.
The coordinates being dual to $\{q_{\,a}\}$ and the components $h_{\,ab}^{\,U}$ 
are  
$$
q_{\,a}^{\,*}
=\frac{\dr\, \phi_{\,\beta}}{\dr q^{\,a}}
=\left[\,(\,q^{\,a}\,)^2\,\right]^{\,\beta-1/2},\qquad\mbox{and}\qquad 
h_{\,ab}^{\,U}
=(2\beta-1)\,\left[\,(\,q^{\,a}\,)^{\,2}\,\right]^{\,\beta-1}\,
\delta_{\,ab},  
\quad\mbox{(no sum)}.
$$
The Legendre transform of $\phi$ has been 
obtained as \fr{Legendre-transform-power-function-with-beta}, 
from which 
$$
q^{\,a}
=\frac{\dr\,\phi^{\,*}}{\dr q_{\,a}^{\,*}}
=
\left(\,(\,q_{\,a}^{\,*}\,)^{\,2}\,\right)^{\,\beta^{\,*}-1/2},\qquad 
h_{\,U}^{\,ab}
=(\,2\beta^{\,*}-1\,)\,\left(\,(\,q_{\,a}^{\,*}\,)^{\,2}\right)^{\,\beta^{\,*}-1}\,
\delta^{\,ab}, 
\quad\mbox{(no sum)},
$$
and 
$$
C_{\,U^{\,*}}^{\,abc}
=\left\{
\begin{array}{cc}
(\,2\beta^{\,*}-1\,)\,(\,2\beta^{\,*}-2\,)\,
\left(\,(\,q_{\,a}^{\,*}\,)^{\,2}\right)^{\,\beta^{\,*}-3/2}
&\mbox{for $a=b=c$}\\
0&\mbox{otherwise}\\
\end{array}
\right..
$$
The components of the Riemannian metric tensor field $h_K$ have been 
calculated as \fr{Toda-h-K-ab}.
\end{Example}
\begin{Example}
In the example below, general interaction potential function is focused, 
where such potential functions have been discussed in the development of 
information geometry. 

To this end, one introduces a positive monotonically increasing  
function and 
considers a generalized logarithmic and exponential 
functions\,\cite{Naudts2002PhysA,OharaWada2010JPhysA}. 
Let $\varphi$ be a function of $\zeta\in\mbbR$ such that 
$$ 
\varphi(\zeta)>0,\qquad\mbox{and}\qquad 
\frac{\dr\,\varphi}{\dr \zeta}>0,\qquad \mbox{for}\quad\zeta>0.
$$
Then define the generalized logarithmic function associated with $\varphi$
$$
\ln_{\,\varphi}(\zeta)
=\int_{\,1}^{\,\zeta}\frac{1}{\varphi(\zeta^{\,\prime})}\dr\zeta^{\,\prime},
$$
and the generalized exponential function as the inverse function of 
$\ln_{\,\varphi}(\zeta)$, so that 
$$
\exp_{\,\varphi}(\,\ln_{\varphi}(\zeta)\,)
=\zeta,\qquad\mbox{and}\qquad
\ln_{\,\varphi}(\,\exp_{\varphi}(\zeta)\,)
=\zeta.
$$
It follows from 
$$
\frac{\dr}{\dr\zeta}\exp_{\,\varphi}(\,\zeta\,)
=\varphi(\,\exp_{\varphi}(\,\zeta)\,)>0,\qquad \mbox{and}\qquad 
\frac{\dr^{\,2}}{\dr\zeta^{\,2}}\exp_{\,\varphi}(\zeta)
=\varphi(\,\exp_{\varphi}(\,\zeta\,)\,)\,
\left.\frac{\dr\,\varphi}{\dr \zeta^{\,\prime}}
\right|_{\,\zeta^{\,\prime}=\exp_{\,\varphi}\zeta}
>0
$$
that the generalized exponential function is strictly convex. 

Choose $\varphi$ and the interaction potential energy function 
$$
\phi_{\,\varphi}(q)
=\int_{\,0}^{\,q}\exp_{\,\varphi}(\,q^{\,\prime}\,)\,\dr q^{\,\prime}.
$$
The Legendre transform of this is obtained as 
\beq
\phi_{\,\varphi}^{\,*}(q^{\,*})
=\int_{\,1}^{\,q^{\,*}}\ln_{\,\varphi}(\,q^{\,*\,\prime}\,)
\,\dr q^{\,*\,\prime},
\label{Legendre-transform-generalized-exponential-function} 
\eeq
from which 
\beq
\frac{\dr}{\dr q^{\,*}}\phi_{\,\varphi}^{\,*}(q^{\,*})
=\ln_{\,\varphi}(\,q^{\,*}\,),\qquad\mbox{and}\qquad 
\frac{\dr^{2}}{\dr q^{\,*\,2}}\phi_{\,\varphi}^{\,*}(q^{\,*})
=\frac{1}{\varphi(\,q^{\,*}\,)}>0.
\label{Legendre-transform-generalized-exponential-function-derivative}
\eeq
With \fr{Legendre-transform-generalized-exponential-function-derivative} 
and  
\fr{Toda-lattice-generalized},  
the explicit form of the dual lattice system is obtained as 
$$
-\,m\,\frac{\dr}{\dr t}\ln_{\,\varphi}(\,-\,\dot{p}_{\,a}\,)
=p_{\,a+1}+p_{\,a-1}-2\,p_{\,a},
$$
or equivalently, 
$$
\frac{m}{\varphi(\,-\,\dot{p}_{\,a}\,)}\frac{\dr^{\,2}p_{\,a}}{\dr t^{\,2}}
=p_{\,a+1}+p_{\,a-1}-2\,p_{\,a},\qquad\mbox{(no sum)}.
$$

Consider the case where the Hessian-information geometric formulation can
be applied to this system. Then the quantities used in
 \fr{canonical-equations-general-second-order-cubic-form} 
are shown below.
The coordinates being dual to 
$\{q^{\,a}\}$  
and the components $h_{\,ab}^{\,U}$ 
are  
$$
q_{\,a}^{\,*}
=\frac{\dr\, \phi_{\,\varphi}}{\dr q^{\,a}}
=\exp_{\,\varphi}(\,q^{\,a}\,),\qquad\mbox{and}\qquad 
h_{\,ab}^{\,U}
=\varphi(\,\exp_{\,\varphi} q^{\,a}\,)\,
\delta_{\,ab},  \quad\mbox{(no sum)}.
$$
The Legendre transform of $\phi_{\,\varphi}$ has been 
obtained as \fr{Legendre-transform-generalized-exponential-function},  
from which 
$$
q^{\,a}
=\frac{\dr\,\phi_{\,\varphi}^{\,*}}{\dr q_{\,a}^{\,*}}
=\ln_{\,\varphi}(\,q_{\,a}^{\,*}\,),\qquad 
h_{\,U}^{\,ab}
=\frac{1}{\varphi\,(\,q_{\,a}^{\,*}\,)}\,\delta^{\,ab},
\quad\mbox{(no sum)},
$$
and 
$$
C_{\,U^{\,*}}^{\,abc}
=\left\{
\begin{array}{cc}

-\,\left[\,\varphi(\,q_{\,a}^{\,*}\,)\,\right]^{\,-2}
(\,\dr\varphi/\dr q_{\,a}^{\,*}\,)&\mbox{for $a=b=c$}\\
0&\mbox{otherwise}\\
\end{array}
\right..
$$
The components of the Riemannian metric tensor field $h_K$ have been 
calculated as \fr{Toda-h-K-ab}.
\end{Example}
\subsection{Vanishing potential systems}
In this subsection it is assumed that 
\begin{itemize}
\item
the dimension of the manifold $\cM$ is $n$, and its 
local coordinates are $p=\{\,p_{\,a}\,\}$, 
\item
a system is a natural Hamiltonian system whose Hamiltonian 
is a strictly convex kinetic energy function, $H=K$ with $K$ being 
a function of $p$. 
\end{itemize}
From these assumptions,     
the condition 
$(\partial^{\,2}K/\partial p_{\,a}\partial p_{\,b})\succ 0$  
is satisfied. 

The canonical equations of motion \fr{canonical-equations-general} are 
$$
\frac{\dr}{\dr t}q^{\,a}
=\frac{\partial K}{\partial p_{\,a}},\qquad\mbox{and}\qquad 
\frac{\dr}{\dr t}p_{\,a}
=0,
$$
from which $\ddot{q}^{\,a}=0$. In these coordinates, one  has
$$
q^{\,a}(t)
=p_{\,*}^{\,a}\,t+q^{\,a}(0),
$$
where $\{p_{\,*}^{\,a}\}$ has been defined as the set of the dual coordinates 
( see definition\,\ref{definition-dual-coordinates-natural-Hamiltonian} ), 
$$
p_{\,*}^{\,a}
=\frac{\partial K}{\partial p_{\,a}}.
$$
One can write the canonical equations as   
$$
\frac{\dr}{\dr t}p_{\,*}^{\,a}
=0, \qquad\mbox{from which}\qquad 
p_{\,*}^{\,a}(t)
=p_{\,*}^{\,a}(0).
$$

Similar to discussions in 
section\,\ref{section-non-vanishing-potential-systems}, one has that 
the triplet $(\cM,\nabla^{\,K},h^{\,K})$ is a Hessian manifold, where 
$\nabla^{\,K}$-affine coordinates are $\{p_{\,a}\}$, $h^{K}$ 
and $\nabla^{\,K}$ have been defined in \fr{metric-h} and  
\fr{definition-connections-K-U}, respectively. Also, 
the triplet $(\cM,\nabla^{\,K^{*}},h^{\,K})$ is a Hessian manifold, where 
$\nabla^{\,K^{*}}$-affine coordinates are $\{p_{\,*}^{\,a}\}$, 
and $\nabla^{\,K^{*}}$ has been defined in \fr{definition-dual-connections-K-U}. 

One immediately has the following.
\begin{Proposition}
In addition to that the set of coordinates $\{p_{\,a}\}$ is constant, 
the set of dual coordinates $\{p_{\,*}^{\,a}\}$ is also constant.
\end{Proposition}

\section{Extension to LC circuit models} 
Since there are some similarities between non-dissipative electric circuit 
models and 
classical Hamiltonian systems, one is interested in 
how to extend the above geometric formulation of Hamiltonian systems to   
circuit theory. In this section it is shown how this is given. 

Consider the following series LC  circuit model 
\beq
\frac{\dr}{\dr t}Q
=I,\qquad 
\frac{\dr}{\dr t}\Phi
=-V,
\label{LC-series-circuit}
\eeq 
where $t\in\mbbR$ is time, $Q\in\mbbR$ 
the electric charge stored in a capacitor 
with electric capacitance $C$,   
$I\in\mbbR$ the current,   
$\Phi\in\mbbR$ the magnetic flux due to $I$
whose flux is stored in an inductor with electric inductance 
$L$,  
$V\in\mbbR$ the capacitor voltage.    
The electromagnetic energy for this circuit model is expressed as  
$$
\cE^{\,*}(V,I)
=\cE_{\mathrm{C}}^{\,*}(V)
+\cE_{\mathrm{L}}^{\,*}(I),\qquad
\cE_{\mathrm{C}}^{\,*}(V)
=\int Q(V)\,\dr V,\qquad 
\cE_{\mathrm{L}}^{\,*}(I)
=\int \Phi(I)\,\dr I,
$$  
where $\cE^{\,*}:\mbbR^2\to\mbbR$ is referred to as a co-energy function, and 
the total Legendre transform of $\cE^{\,*}$ is the energy function denoted by 
$\cE$. 
The value $\cE_{\mathrm{C}}^{\,*}(V)$  
is interpreted as the energy due to  
the capacitor, and $\cE_{\mathrm{L}}^{\,*}(I)$ the energy due to the 
inductor.  
The constitutive relations  are 
\beq
Q(V)
=\frac{\partial \cE^{\,*}}{\partial V}
=\frac{\dr \cE_{\mathrm{C}}^{\,*}}{\dr V},\qquad\mbox{and}\qquad 
\Phi(I)=
\frac{\partial \cE^{\,*}}{\partial I}
=\frac{\dr \cE_{\,\mathrm{L}}^{\,*}}{\dr I}.
\label{circuit-constitutive-relations}
\eeq
In what follows it is assumed that 
\begin{itemize}
\item 
$\cE_{\,\mathrm{C}}^{\,*}$ and $\cE_{\,\mathrm{L}}^{\,*}$ 
 are strictly convex :  
$$
\frac{\dr^2\cE_{\,\mathrm{C}}^{\,*}}{\dr V^2}>0,\qquad\mbox{and}\qquad 
\frac{\dr^2\cE_{\,\mathrm{L}}^{\,*}}{\dr I^2}>0,
$$
in some domains.
\end{itemize}
Similar to definition\,\ref{definition-dual-coordinates-natural-Hamiltonian},
one defines the following.
\begin{Def}
{\bf (Dual coordinates).}\ The coordinates $Q$ and $\Phi$ defined by 
\fr{circuit-constitutive-relations} are referred to as dual coordinates. 
In particular, $Q$ is referred to as being dual to $V$, and $\Phi$ is 
referred to as being dual to $I$. 
\end{Def}
From the assumption it can be shown that 
$$
V=\frac{\partial \cE}{\partial Q}
=\frac{\dr \cE_{\mathrm{C}}}{\dr Q},\qquad \mbox{and}\qquad 
I=\frac{\partial \cE}{\partial \Phi}
=\frac{\dr \cE_{\mathrm{L}}}{\dr \Phi},
$$
where $\cE_{\mathrm{C}}$ is the Legendre transform of $\cE_{\mathrm{C}}^{\,*}$, 
and $\cE_{\mathrm{L}}$ the Legendre transform of $\cE_{\mathrm{L}}^{\,*}$.
 
The series LC circuit model \fr{LC-series-circuit} can then be written as 
$$
\frac{\dr^2\,Q}{\dr t^{\,2}}
=-\,\frac{\dr^2\cE_{\,\mathrm{L}} }{\dr \Phi^2}\,V,
$$
from which 
\beq
\frac{\dr^2}{\dr t^2}\left(\,
\frac{\dr\cE_{\mathrm{C}}^{\,*}}{\dr V}
\,\right)
=-\,\frac{\dr^2 \cE_{\,\mathrm{L}}}{\dr \Phi^2}V.
\label{LC-series-equation-general-second-order}
\eeq
This equation is an analogue of 
\fr{canonical-equations-general-second-order}. 

The following is analogous to   
theorem\, \ref{fact-generalized-Toda-dual-transformed-equations}.
\begin{Thm}
\label{LC-circuit-Toda-dual-transformed-equations}
{\bf (Generalized dual transformed equations for LC circuits ).}  
Assume that $\dr^{\,2}\cE_{\,\mathrm{L}}/\dr \Phi^{\,2}$ 
 is constant. Then  
the series LC circuit model is written of the form 
$$
\frac{\dr}{\dr t}\left(\left.
\frac{\dr \cE_{\mathrm{C}}^{\,*}}{\dr V}\right|_{V=-\dot{\Phi}}\right)
=\frac{\dr^2 \cE_{\,\mathrm{L}}}{\dr \Phi^2}\Phi+\const.
$$
\end{Thm}
\begin{Proof}
A way to prove this is analogous to that of 
theorem\, \ref{fact-generalized-Toda-dual-transformed-equations}.
\qed
\end{Proof}

One has the following theorem, and this is 
analogous to theorem\, 
\ref{fact-canonical-equation-Hessian-language}.
\begin{Thm}
\label{LC-circuit-Hessian-language}
{\bf (LC circuits written in terms of Hessian geometry).}\  
The series LC circuit model in terms of the dual coordinates 
can be written in the form 
$$
\check{C}^{\,*}\,(\dot{V})^{\,2}\,
+h_{\,\mathrm{C}}^{\,*}\ddot{V}
=-\,h_{\,\mathrm{L}}V,\qquad 
\check{C}^{\,*}
:=\frac{\dr^3\cE_{\,\mathrm{C}}^{\,*}}{\dr V^3},\quad 
h_{\,\mathrm{C}}^{\,*}
:=\frac{\dr^2\cE_{\,\mathrm{C}}^{\,*}}{\dr V^2},\quad
h_{\,\mathrm{L}}
:=\frac{\dr^2\cE_{\,\mathrm{L}}}{\dr \Phi^2}.
$$
\end{Thm}
\begin{Proof}
A way to prove this is analogous to that of 
theorem\,\ref{fact-canonical-equation-Hessian-language}.
\qed
\end{Proof}
\begin{Remark}
It is straightforward to see that $\check{C}^{\,*}$ is the component of a  
cubic form.
\end{Remark}

Note that constitutive relations are obtained by differentiating energy 
or co-energy functions as in \fr{circuit-constitutive-relations}. 
By contrast, if  constitutive relations 
are expressed as monotonically increasing functions, 
then integration of such constitutive relations 
gives strictly convex 
energy or co-energy functions.

The following is an example. Unlike the general discussion above, 
constitutive relations are given first. 
Second, co-energy and energy functions are calculated. Then the 
second order equation written in terms of introduced geometric objects
is shown.   
\begin{Example}
Choose the following constitutive relations 
$$
\Phi(I)
=LI,\qquad\mbox{and}\qquad 
Q(V)
=Q_{\,0}\ln\left(1+\frac{V}{V_{\,0}}\right),
$$
where $L>0$, and  
$Q_{\,0},V_{\,0}>0$ are constants. 
Note that 
$$
Q_{\,0}\neq \lim_{V\to 0}Q(V),\qquad\mbox{ and }\qquad
\lim_{V\to 0}\frac{\dr Q}{\dr V}
=\frac{Q_{\,0}}{V_{\,0}}.
$$
Then the co-energy function is  
$$
\cE^{\,*}
(V,I)
=\cE_{\,\mathrm{L}}^{\,*}(I)+\cE_{\,\mathrm{C}}^{\,*}(V),
$$
where 
$$
\cE_{\,\mathrm{L}}^{\,*}(I)
=L\frac{I^{\,2}}{2},\qquad\mbox{and}\qquad
\cE_{\,\mathrm{C}}^{\,*}(V)
=Q_{\,0}V_{\,0}
\left[\,\left(1+\frac{V}{V_{\,0}}\right)\, 
\ln\left(1+\frac{V}{V_{\,0}}\right)
-\frac{V}{V_{\,0}}\,\right].
$$
They are strictly convex for $\{(V,I)\,|\,V+V_{\,0}>0\}$, since 
$$
\frac{\dr^2\,\cE_{\,\mathrm{L}}^{\,*}(I)}{\dr I^{\,2}}
=L>0,\qquad\mbox{and}\qquad
\frac{\dr^2\,\cE_{\,\mathrm{C}}^{\,*}(V)}{\dr V^{\,2}}
=\frac{Q_{\,0}}{V+V_{\,0}}>0.
$$
The total Legendre transforms of $\cE_{\,\mathrm{L}}^{\,*}$
and $\cE_{\,\mathrm{C}}^{\,*}$,   
$$
\cE_{\,\mathrm{L}}(\Phi)
=\sup_{I}\left[\,\Phi I-\cE_{\,\mathrm{L}}^{\,*}(I)\,\right],\qquad 
\mbox{and}\qquad 
\cE_{\,\mathrm{C}}(Q)
=\sup_{V}\left[\,Q V-\cE_{\,\mathrm{C}}^{\,*}(V)\,\right],
$$
are obtained from the relation $1+V/V_{\,0}=\exp(Q/Q_{\,0})$ as 
$$
\cE_{\,\mathrm{L}}(\Phi)
=\frac{\Phi^{\,2}}{2\,L},\qquad \mbox{and}\qquad 
\cE_{\,\mathrm{C}}(Q)
=V_{\,0}\left(\,Q_{\,0}\,\e^{\,Q/Q_{\,0}}-Q-Q_{\,0}\,\right).
$$
From $V(Q)=V_{\,0}(\exp(Q/Q_{\,0})-1)$, one verifies that 
$$
V_{\,0}\neq\lim_{Q\to 0} V(Q),\qquad\mbox{and}\qquad 
\lim_{Q\to 0}\frac{\dr V}{\dr Q}
=\frac{V_{\,0}}{Q_{\,0}}.
$$
The well-known quadratic co-energy and energy functions are  
obtained with the Taylor expansion of $\cE_{\mathrm{C}}^{\,*}(V)$ and  that of 
$\cE_{\,\mathrm{C}}(Q)$ as 
$$
\cE_{\mathrm{C}}^{\,*}(V)
=\frac{C_{\,0} V^{\,2}}{2}+\mbox{(higher order terms)},\qquad\mbox{and}\qquad 
\cE_{\,\mathrm{C}}(Q)
=\frac{Q^{\,2}}{2\,C_{\,0}}+\mbox{(higher order terms)},
$$
where 
$C_{\,0}:=Q_{\,0}/V_{\,0}$.

The second order circuit equation in terms of the dual coordinate $V$ is 
obtained from \fr{LC-series-equation-general-second-order} as  
$$
\frac{\dr^2}{\dr t^2}\left[\,Q_{\,0}\ln\left(1+\frac{V}{V_{\,0}}\right)\,
\right]
=-\, \frac{1}{L}V.
$$ 
The generalized dual transformed equation is 
obtained from 
theorem\,\ref{LC-circuit-Toda-dual-transformed-equations}
as 
$$
\frac{\dr}{\dr t}\left[\,Q_{\,0}\ln\left(1-\frac{1}{V_{\,0}}\frac{\dr\,\Phi}{\dr t}\right)\,
\right]
=\frac{1}{L}\Phi+\const.
$$
 
The components of geometric objects in 
theorem\,\ref{LC-circuit-Hessian-language} are calculated as 
$$
h_{\,\mathrm{C}}^{\,*}
=\frac{Q_{\,0}}{V+V_{\,0}},\qquad
\check{C}^{\,*}
=\frac{\dr h_{\,\mathrm{C}}^{\,*}}{\dr V}
=-\,\frac{Q_{\,0}}{(V+V_{\,0})^{\,2}},\qquad
\mbox{and}\qquad
h_{\,\mathrm{L}}
=\frac{1}{L}.
$$

\end{Example}
\section{Conclusions}
This paper has offered a formulation of a class of classical Hamiltonian 
systems in terms of Hessian-information geometry.
It has been shown that the $\alpha$-connection, cubic forms and so on, 
invented in information geometry,  appear  
in canonical equations of motion. 
From the theorems in this paper, 
one significance of the proposed geometric 
formulation of 
Hamiltonian systems is to give an explicit form of 
Toda's dual transformed equations from a given lattice system.
In this sense, dual in the sense of Toda is equivalent to that in the 
sense of Legendre. With this formulation, some explicit forms of 
dual lattice systems have been obtained. 
Also this paper has offered a similar formulation of non-dissipative 
electric circuit models.

There are some potential future works that follow from this paper. 
One is to apply the present approach to dissipative systems 
written in terms of 
contact geometry\,\cite{Bravetti2017Annual,Bravetti2017entropy}.
Since the present study has been restricted 
to conservative Hamiltonian systems,  
it is interesting to see if this approach can be extended for   
dissipative systems.    
With these future works based on this study, 
it is expected that the proposed geometric formulation yields  
a step to importing various theorems in Hessian-information geometry to
 Hamiltonian mechanics and electric circuit theory, and vice versa.

\section*{Acknowledgments}
The first author (SG) 
was partially supported by 
Toyota Physical and Chemical Research Institute in Japan. 
The second author (TW) {was} partially supported by Japan Society for the 
Promotion of Science (JSPS) Grants-in-Aid for Scientific Research (KAKENHI)
Grant Number JP17K05341.
The authors would like to thank M Koga at Nagoya University for 
giving various comments on this paper. 

\appendix
\section{A brief explanation of information and Hessian geometries}
This appendix provides a brief explanation of information geometry and 
Hessian geometry. This is intended to give a rough sketch of these geometries  
for the readers who are not familiar with them but familiar with Riemannian 
geometry.
For more comprehensive discussions for these geometries see 
\cite{AN} and \cite{Ay2017} from a mathematical viewpoint, and 
\cite{Wagenaar1998} from a physical viewpoint.       

Information geometry is a geometrization of mathematical statistics. 
In particular, most of cases parametric distribution functions are focused. 
In such a case, a finite set of parameters associated 
with a given probability distribution function
is identified with a set of coordinates of a manifold.  
Although a metric tensor field called the Fisher metric tensor field 
is often introduced, 
information geometry is not exactly classified as a Riemannian geometry.
Instead, a manifold, a metric tensor field, and 
some two connections play roles
in information geometry. 
The following is a list of geometrical objects used in information geometry.
\begin{itemize}
\item
Coordinates: parameters for expressing a distribution function
\item
Components of a metric tensor field: 
the Fisher information matrix 
\item
Two connections: a connection $\nabla$ and its dual connection $\nabla^{\,*}$, 
which are torsion-free 
and need not to be the Levi-Civita connection. Here dual connection 
is defined through the Fisher metric tensor field $g$ so that  
$X[g(Y,Z)]=g(\nabla_XY,Z)+g(Y,\nabla_X^{\,*}Z)$ is satisfied 
for any vector fields $X,Y$ and $Z$.  
\end{itemize}
The importance of these objects,  or a structure,  
is summarized as the Chentsov theorem.  
A manifold equipped with this structure is referred to as a statistical
manifold.

Dually flat spaces are particularly important in information geometry.
If a connection $\nabla$ and dual connection $\nabla^{\,*}$ 
are flat, then such a statistical manifold 
is referred to as a dually flat space, and is denoted by  
$(\cM,g,\nabla,\nabla^{\,*})$, where $\cM$ is a manifold 
expressing probability distribution functions.
As briefly explained below, a dually flat space is related to  
the Legendre transform and dual coordinates.  
In a dually flat space, by definition, $\nabla$-affine coordinates and 
$\nabla^{\,*}$-affine coordinates exist. 
Let $\theta=\{\theta^{\,a}\}$ and $\eta=\{\eta_{\,a}\}$ be such coordinates. 
It turns out that strictly convex functions 
are guaranteed to exist.  
With these functions, denoted by $\psi$ and $\varphi$, 
the coordinate transforms between $\theta$ and $\eta$
can be written as 
 $\eta_{\,a}=\partial\psi/\partial\theta^{\,a}$ and 
$\theta^{\,a}=\partial\varphi/\partial\eta_{\,a}$. In addition, 
the Legendre transform of $\psi$ is $\varphi$, and that of $\varphi$ is $\psi$.
These relations above are employed in the main text. 
In addition, since $\nabla$ and $\nabla^{\,*}$ are flat, geodesic curves are immediately 
obtained as follows.  
Geodesic curves with respect to $\nabla$ reduce to  
$\theta^{\,a}(t)=\dot{\theta}^{\,a}(0)t+\theta^{\,a}(0)$ with $t\in\mbbR$ 
parameterizing a set of points on a geodesic curve.  
Similarly, geodesic curves with respect to $\nabla^{\,*}$ reduce to
$\eta_{\,a}(t)=\dot{\eta}_{\,a}(0)t+\eta_{\,a}(0)$.

Hessian geometry is linked to information geometry 
as mentioned in Introduction of this paper. 
In what follows this is explained. 
As discussed in the main text,   
Hessian geometry is a geometry where a Hessian structure 
$(\check{\nabla},\check{g})$   
is provided, and strictly convex functions play fundamental roles. Here 
$\check{\nabla}$ is a connection, and $\check{g}$ a (pseudo) Riemannian 
metric tensor field. It can be shown that a strictly convex function 
induces a Hessian manifold. 
To show a link between information geometry and Hessian geometry,  
consider the cumulant generating function of a 
distribution function belonging to 
the exponential family. 
Here, the cumulant generating function of the exponential family 
is convex, where the exponential family is 
a class of parametric probability distribution functions. 
It can be shown that a strictly convex function induces a dually flat space.
Thus, by restricting a domain for the cumulant generating function so that 
it is strictly convex, one has a Hessian manifold and a dually flat space. 
This shows how information geometry is related to Hessian geometry.



\begin{thebibliography}{99}
\bibitem{Shima}
Shima H\, 2007
{\it The geometry of Hessian structures}\, (Singapore: World Scientific)  
%
\bibitem{AN} 
Amari S and  Nagaoka H\, 2000   
{\it Methods of information geometry}\, (Oxford: Oxford University Press)
%
\bibitem{Ay2017}
Ay N {\it et al}\ 2017   
{\it Information Geometry}\, (Newyork: Springer)
\bibitem{Matsuzoe2013}
Matsuzoe H and  Henmi M\,2013    
Hessian Structures on Deformed Exponential Families   
 {\it Geometric Science of Information. Lecture Notes in Computer Science}  
{\bf 8085},  ed F Nielsen , F Barbaresco F  
(Berlin:Springer) pp275--282
%
\bibitem{Furuhata2009}
Furuhata H\, 2009 
Hypersurfaces in statistical manifolds   
{\it Differential Geometry and its Applications}  {\bf 27}    
(Amsterdam: Elsevier) pp420--429
%
\bibitem{Arnold}
Arnold V I 1997\,{\it Mathematical Methods of Classical Mechanics}, 2nd edn   
(New York:Springer)

\bibitem{Silva2008}
da Silva A C 2008\,  
{\it Lectures on Symplectic Geometry}, 2nd edn  
(New York: Springer)

%
\bibitem{Goto2016}
Goto S\, 2016 
Contact geometric descriptions of vector fields on dually flat spaces and
their applications in electric circuit models and nonequilibrium statistical
mechanics 
{\it J. Math. Phys.} {\bf 57} 102702

\bibitem{Eberard2006}
Eberard D, Maschke B M  and Van Der Schaft A J\, 2006   
Energy-conserving formulation of RLC-circuits with linear resistors  
{\it Proc. 17th Int.  Symp. on Mathematical
Theory of Networks and Systems}, (Kyoto, Japan), pp71--76



\bibitem{BN2016}
Boumuki N and Noda T\,2016 
On gradient and Hamiltonian flows on even dimensional dually flat spaces
{\it Fun. J. Math. and Math. Sci.} {\bf 6} pp51--66


\bibitem{Goldstein}
  Goldstein H, Pool C P Jr and  Safko J L\, 2014  
{\it Classical Mechanics}, 3rd edn, 
(Harlow:Pearson Education)

\bibitem{Callen}
Callen H B\, 1985 
{\it Thermodynamics and introduction to Thermostatistics}, 2nd edn, 
(New York: Wiley)

\bibitem{Quevedo2008}
Quevedo H\, 2008 
Geometrothermodynamics of black holes   
{\it Gen. Relativ. Gravit.}  {\bf 40} pp971–-984 

\bibitem{Goto2015}
Goto S\, 2015  
Legendre submanifolds in contact manifolds as attractors and geometric
nonequilibrium thermodynamics  
{\it J. Math. Phys.} {\bf 56} 73301 
 
\bibitem{Wada2015}
Wada T, Matsuzoe H and Scarfone A M \, 2015
Dualistic Hessian Structures Among Thermodynamic Potentials in the 
$\kappa$-Thermostatistcis 
{\it Entropy} {\bf 17} pp7213--7229

\bibitem{Teruel2013}
Teruel G R P\,2013  
An alternative formulation of Classical Mechanics based on an analogy with 
Thermodynamics  
{\it Eur. J. Phys.} {\bf 34} pp1589--1599

\bibitem{LZ2017}
Leok M and Zhang J\,2017  
Connecting Information Geometry and Geometric Mechanics  
{\it Entropy} {\bf 19} 518  


\bibitem{Toda1967}
Toda M\,1967\, 
Vibration of a Chain with Nonlinear Interaction   
{\it J. Phys. Soc. Jpn.} {\bf 22} pp431--436
  
\bibitem{Fujiwara1995}
Fujiwara A and Amari S\,1995 
Gradient systems in view of information geometry  
{\it Physica D} {\bf 80} pp317--327


\bibitem{Nakamura1994}
Nakamura Y\,1994  
A tau-function of the finite nonperiodic Toda lattice  
{\it Phys. Lett. A} {\bf 195} pp346--350

\bibitem{Miwa2000}
Miwa T, Jimbo M  and  Date E, 2000 
{\it Solitons: Differential Equations, Symmetries and Infinite Dimensional 
Algebras} (Cambridge tracts in Mathematics)
(Cambridge:Cambridge University Press)
 
\bibitem{Naudts2002PhysA}
Naudts J\, 2002
Deformed exponentials and logarithms in generalized thermodynamics   
{\it Physica A} {\bf 316} pp323--334

\bibitem{OharaWada2010JPhysA}
Ohara A and Wada T\, 2010
Information geometry of q-Gaussian densities and
behaviors of solutions to related diffusion equations  
{\it J. Phys. A:Math.Theor.} {\bf 43} 035002


\bibitem{Bravetti2017Annual}
Bravetti A,  Cruz H and Tapias D\, 2017
Contact Hamiltonian mechanics  
{\it Ann. Phys.} {\bf 376}  pp17--39

\bibitem{Bravetti2017entropy}
Bravetti A\, 2017
Contact Hamiltonian Dynamics: the Concept and its use  
{\it Entropy} {\bf 19}  535

\bibitem{Wagenaar1998} 
Wagenaar D\,1998  
Information geometry for neural networks
{\it MSc Thesis} King's college London 
(www.danielwagenaar.net/papers/98-Wage2.pdf) 

 
\end{thebibliography}
\end{document}